\documentclass[prb,preprint,titlepage,showpacs,preprintnumbers,amsmath,amssymb,floats]{revtex4-2}
\usepackage{blindtext}
\usepackage{graphicx}
\usepackage{amsmath}
\usepackage{amssymb}
\usepackage{color}

\newcommand{\be}{\begin{eqnarray}}
\newcommand{\ee}{\end{eqnarray}}
\newcommand{\beq}{\begin{equation}}
\newcommand{\eeq}{\end{equation}}
\usepackage[toc,page]{appendix}

\begin{document}
\title{Quantum-criticality and superconductivity in \\ twisted  transition metal di-chalcogenides}
\author{A.V. Chubukov}
\affiliation{Department of Physics, University of Minnesota, Minneapolis, MN. 55455 }
\author{C. M. Varma}
\affiliation{Visiting Scholar, Department of Physics, University of California, Berkeley, CA. 94720\\
Department of Physics, University of California, Riverside, CA. 92521 }
\thanks{ Emeritus}
\date{\today}
\begin{abstract}
We analyze a model for electronic structure and interactions in twisted transition metal chalcogenide WSe$_2$ for
superconductivity. In this material,
 spin-orbit scattering locks the z-components of spins of low-energy fermions near the Dirac ${\bf K}$ and ${\bf K}'$ points of the hexagonal Brillouin zone, reducing the symmetry of spin-spin interactions to that of an xy model.
 We show that a nominally repulsive 4-fermion interaction gives rise to an attraction for pairing in a two-component
   $E^{-}$ channel, which is a hexagonal lattice representation of
  $\ell =1$ channels.
  The gap function is inversion-odd and
 a linear combination of  spin singlet and spin triplet.  At weak coupling superconductivity emerges via the Kohn-Luttinger mechanism; we compute $T_c$ for the Fermi-level lying close to the van Hove singularity.  At strong coupling, the pairing is mediated by XY magnetic  fluctuations peaked at momenta ${\bf K}-{\bf K}' = 2 {\bf K}$ and we estimate $T_c$  using the form of  the form of the quantum-critical XY fluctuations, displaying $\omega/T$ scaling.
     \end{abstract}
     \maketitle

\section{Introduction}
The fabrication of twisted multilayer Graphene (TBG) \cite{Cao:2018xq, Cao:2018lk, EAndreiREV2021}
and twisted transition metal di-chalcogenides (t-TMD) \cite{Pasupathy2021, Mak_WSE2_2022} heterostructures with tunable gates allows electron densities in the bands close to the chemical potential to be continuously varied as well as the ratio of electronic interaction energies  to the kinetic energy.  Of special interest are regions of the phase diagram in which the insulator to metal transition is accompanied by a characteristic fan-like quantum-critical region in which the resistivity is linear in temperature, followed at lower temperature with a superconducting region \cite{Cao:2018xq, Efetov2019}. On suppressing superconductivity with a magnetic field $H$, a region of resistivity linear in $H$ is also found in TBG \cite{Efetov2019}.
Recently,
 very similar quantum-critical properties as well as superconductivity have been found in twisted bilayer TMD (tTMD) 
 WSe$_2$ 
 in two different experiments with twist angles near $3^\circ-$ (Ref. \cite{Mak_supercond_2024}) and near $5^\circ$ (Ref. \cite{Pasupathy2024superconductivity}). These include the variation of resistivity linear in $T$ and applied field $H$.
 In both systems, superconductivity is enhanced by applying a finite displacement field $D$.

In this work we focus on superconductivity, keeping in mind that, quite generally, the symmetry and the parameters of superconductivity
 come from the interaction of fermions with the same fluctuations which determine the normal state properties.

 We start with a microscopic low-energy model of tTMD in a finite displacement field proposed in earlier works\cite{MacDonald_TMD2018,DasSarma_TMD2020,Millis_TMD2021,Devakul:2021kn,Scheurer2024,*Scheurer2024_1,Guerci2024}.
 The model describes two topmost bands with relevant momenta near ${\bf K}$ and ${\bf K}'$ points in the moire Brillouin zone.
 The bands are approximately spin degenerate at zero displacement field D, despite strong spin-orbit interaction and spin-valley locking within each layer. At a finite $D$, this degeneracy is, however, lifted and excitations near ${\bf K}$ and ${\bf K}'$ are spin-polarized in opposite directions in a plane perpendicular to the layers.

We first analyze superconductivity at weak coupling  and  then extend the results to moderate/strong coupling.
For a weak coupling analysis, we assume for definiteness that the system is close to the line in the $(D, \nu)$ plane, with $\nu$ the filling factor, where the van Hove singularity in the band dispersion coincides with the Fermi level at ${\bf K}$ for $\uparrow$ spin and at ${\bf K}'$ for $\downarrow$ spin.  We approximate the inter-valley and intra-valley density-density interactions for fermions near ${\bf K}$  and ${\bf K}'$ by $U$,
 and  assume that the coherence factors associated with the projection onto the topmost valence band are non-singular near ${\bf K}$ and ${\bf K}'$   so that their effects can be incorporated into $U$.
  We then obtain the pairing interaction to order $U^2$, implementing the Kohn-Luttinger mechanism,  and solve the linearized gap equation.  We find that the effective pairing interaction at order $U^2$ is attractive in two-component $E^{-}$ and one-component $A^{-}$ channels  (lattice versions of $p-$wave and $f-$wave).
  The attraction comes from the "crossed" diagram, while the RPA-type "bubble" diagrams cancel out with the vertex correction diagrams. Attraction in $E^{-}$ is about twice stronger than in $A^{-}$.  We solved for $T_c$ in the $E^{-}$ channel and find the explicit expression for  $T_c$. Because of the Hove singularity, $T_c$  has a power-law rather than exponential dependence on $U$: $T_c \propto U^3$.  The spin-structure of the gap function is $(\uparrow \downarrow)$, which is a mixture of a spin-singlet and a spin-triplet.  The momentum dependence is highly non-monotonic: the gap $\Delta ({\bf k})$ scales linearly in $k$ at small $k$ and as $1/k^2$ at large $k$.  The angular dependence of $\Delta ({\bf k})$ also varies between  the two limits.

 We  next  show that the effective interaction $\Gamma$  to order $U^2$ from the crossed diagram can be equally viewed as a magnetic interaction with momentum transfer near $2{\bf K}$.
 We show how this happens even though the only the small momentum part of the interactions $U$ within a given valley is involved. We extend this interaction to higher orders in $U$
  to obtain the dressed RPA XY magnetic susceptibility $\chi (Q)$, and argue that in the parameter range, relevant for experiments,  in which the system is near the onset of Antiferromagnetic  XY magnetic order,  the pairing can be viewed as mediated by fluctuations of an XY AFM order parameter.  We conjecture that the gap symmetry remains the same as at weak coupling, but the expression for $T_c$ changes. In particular, dynamical aspects become relevant, and the gap equation becomes an integral equation not only in momentum, but also in frequency.

   For an estimate of $T_c$ in this case, we use the form of the dynamical  XY AFM susceptibility,  obtained earlier by renormalization group methods and
 quantum-Monte-Carlo calculations~\cite{Aji-V-qcf1,Varma_IOPrev2016}. This reproduces microscopically the Marginal Fermi liquid paradigm,
 required at a phenomenological level~\cite{CMV-MFL} to explain the variety of experiments
 in the  cuprates
\cite{Kaminski-diARPES, simon-cmv, Bourges-rev}, in heavy-fermion materials near their antiferromagnetic (AFM) quantum-critical point
(AFM-QCP) \cite{Lohneysen1996}, as well as  in Fe-based metals \cite{ShibauchiQCP} also near their AFM-QCP.
 That the same XY AFM model emerges naturally in tTMD, which displays similar quantum-critical behavior in the normal state suggests that it well may be a universal model for  different microscopic realizations of Marginal Fermi liquid behavior.  (Other theoretical models have also been proposed~
 \cite{Sachdev_2011,VNS,TailleferREV2010,klein2020,palee2021,Cano2022,Sachdev_science2023}.)  We show from general symmetry considerations that  momentum dependence of such fluctuations determines the pairing symmetry, which reproduces the $E^{-}$ symmetry obtained microscopically within the Kohn-Luttinger scenario,
  while their frequency dependence determines $T_c$ differently.

 Our results about $E^{-}$ gap symmetry and an attraction in the subleading $A^{-}$ channel agree with two recent studies of superconductivity in tTMG~\cite{Scheurer2024,*Scheurer2024_1,Guerci2024}.  We comment on the relation of our results to these works later in the paper.

The structure of the paper is the following.
   In Sec. (\ref{Sec1}), we introduce the model and discuss the electronic structure, Van Hove singularity, and possible charge and spin order parameters.
  In Sec. (\ref{Sec2}), we investigate the superconductivity
  symmetry and superconducting transition temperature in weak coupling, using the  Kohn-Luttinger  mechanism.
   In Sec. (\ref{Sec3}), we extend the analysis to the parameter range where the system is about to develop an XY AFM order and discuss the pairing mediated by XY AFM fluctuations.
We present some concluding remarks in Sec. (\ref{Sec4}).

\section{Electronic structure and possible
 order \\ parameters in t-TMD's}
\label{Sec1}
\begin{figure}
 \includegraphics[width= 0.5\columnwidth]{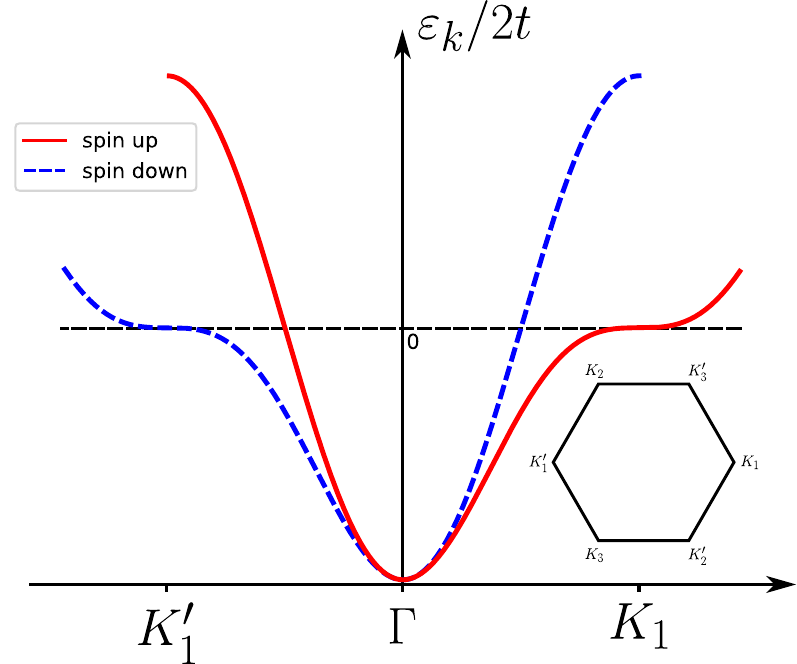}
\caption{The Brillouin zone and the electronic dispersion. The points
${\bf K}_{1}, {\bf K}^{'}_1 = (\pm 2\pi/3, \pm 2\pi/\sqrt{3}), {\bf K}_{2}, {\bf K}^{'}_2= (\mp 4\pi/3,0),\pm 2\pi/\sqrt{3}) {\bf K}_{3}, {\bf K}^{'}_3 = (\pm  2\pi/3, \mp 2\pi/\sqrt{3})$, respectively. The distances between ${\bf K}_i$ and ${\bf K}_{j}$ and between ${\bf K}^{'}_i$ and ${\bf K}^{i}_{j}$ are reciprocal lattice vectors.
The dispersion is spin-split and is shown for $\phi = \pi/6$, where there is a VH singularity at ${\bf K}_i$ for spin-$\uparrow$ band and at ${\bf K}^{'}_i$  spin-$\downarrow$ band.}
 \label{fig:1}
\end{figure}

The electronic structure of the hole doped TMD  homobilayers  has been investigated in detail  \cite{MacDonald_TMD2018,DasSarma_TMD2020,Millis_TMD2021,Devakul:2021kn,Scheurer2024,*Scheurer2024_1,Guerci2024} as a function of twist angle and a displacement field across the layers.
  At a finite displacement field their effective Hamiltonian does {\it not} possess $SU(2)$ spin rotational symmetry, but has only $U(1)$ spin symmetry associated with $s_z$ conservation. This is due principally to two effects: Atomic spin-orbit coupling which is strong in a single layer TMD and gives rise to spin-valley locking of the states in valleys near
 ${\bf K}_0$ and ${\bf K}^{'}_0$ in the original Brillouin zone.  After twisting, the points  ${\bf K}_0$ from, e.g., top layer and ${\bf K}^{'}_0$ from the bottom layer merge onto the point ${\bf K}$ in the moire Brillouin zone, and and  ${\bf K}^{'}_0$ from the top layer and ${\bf K}_0$ from the bottom layer merge onto ${\bf K}^{'}$. Second, at low carrier density, inter-valley scattering can be neglected as it requires a large momentum transfer $Q=K -K'\gg 1/a_M$.
  As long as one restricts with the quadratic dispersion near ${\bf K}_0$ and ${\bf K}^{'}_0$ in each layer,  both bonding and antibonding bands in the moire Brillouin zone remain spin-degenerate.  However, in the presence of a displacement field, the degeneracy is broken, and excitations
  become spin-locked. In particular,
    the band energy at the Dirac points in the moire Brillouin zone
     ${\bf K}_{1}, {\bf K}^{'}_1 = (\pm 2\pi/3, \pm 2\pi/\sqrt{3}), {\bf K}_{2}, {\bf K}^{'}_2 = (\mp 4\pi/3,0),\pm 2\pi/\sqrt{3}), {\bf K}_{3}, {\bf K}^{'}_3 = (\pm  2\pi/3, \mp 2\pi/\sqrt{3})$
      satisfies  $E({\bf K}_i, \uparrow) = E ({\bf K}^{'}_i, \downarrow)$, but $E({\bf K}_i, \uparrow) \neq E ({\bf K}_i, \downarrow)$ (see Fig. \ref{fig:1}).
\begin{figure}
 \includegraphics[width= 0.9\columnwidth]{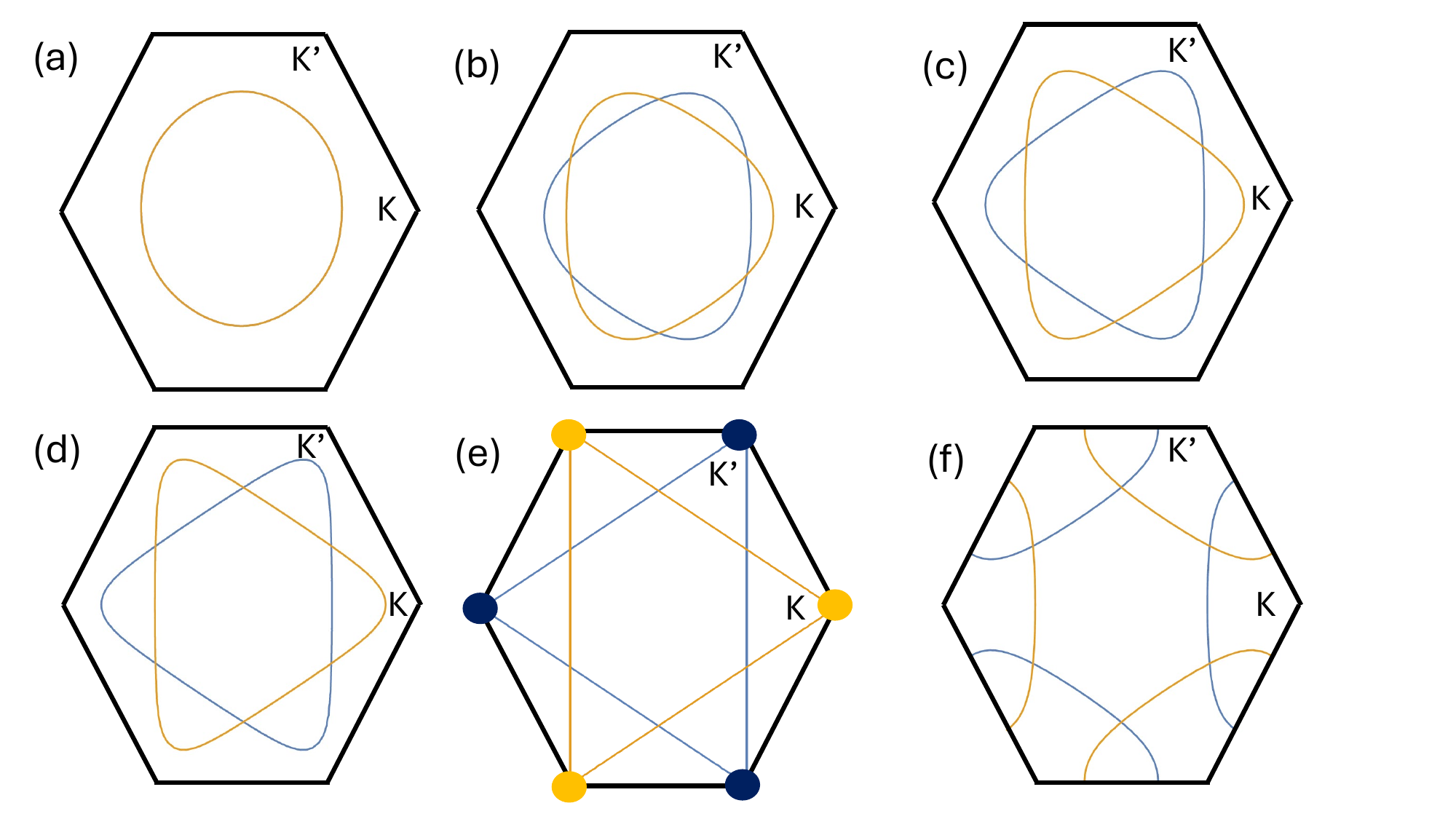}
\caption{The  contours $\epsilon_{\uparrow} ({\bf k})  =0$ (yellow) and
  $\epsilon_{\downarrow} ({\bf k}) =0$ (blue) in the moire Brillouin zone for $\phi =0$ (a), $\pi/10$ (b), $\pi/7$ (c), $\pi/6.5$ (d), $\pi/6$ (e), $\pi/5$ (f).  At $\phi =\pi/6$, the contours touch the Brillouin zone boundary at ${\bf K}_i$ (yellow circles) and ${\bf K}^{'}_i$ ( blue  circles). Expansion near these points yields a cubic dispersion, specific to a higher-order VH singularity.}
 \label{fig:FS}
\end{figure}
In explicit form, the  one-electron band-structure near the top of the valence band in the Moire-lattice for t-TMD is given by \cite{Millis_TMD2021,Devakul:2021kn,Scheurer2024,*Scheurer2024_1,Guerci2024}
\be
\label{bs}
H = -2 |t| \sum_{{\bf K, R_i}, \alpha \sigma} \cos\big({\bf k_{\alpha} \cdot R_i} + \phi \sigma_z\big) c^+_{{\bf k},\alpha, \sigma} c_{{\bf k},\alpha, \sigma}.
 \ee
 Here $R_1 = (1,0), R_2 = (-1/2, \sqrt{3}/2), R_3 = (-1/2, -\sqrt{3}/2)$
 give the position of the atoms on the triangular lattice. $\phi$  represents the displacement field which splits the
  energies of up and  down spins.
  At $\phi =0$,  excitations in each valley are spin degenerate, $\epsilon_{\uparrow} ({\bf k}) = \epsilon_{\downarrow} ({\bf k})$.
   At $\phi \neq 0$, spin degeneracy is broken.  In particular,
 $\epsilon_{\uparrow} (K_i) = \epsilon_{\downarrow} (K'_i) = 6|t| \cos(\phi + \pi/3)$ and
 $\epsilon_{\downarrow} (K_i) = \epsilon_{\uparrow} (K'_i) = 6|t| \cos(\phi - \pi/3)$, i.e., energies remain equal for
  a given spin projection at the $K_i$ points and  opposite spin projection at the $K'_i$ points.
  For definiteness, we assume $\phi >0$.  We show the evolution of the contours $\epsilon_{\uparrow} ({\bf k})  =0$ and
  $\epsilon_{\downarrow} ({\bf k}) =0$ with $\phi$ in Fig. \ref{fig:FS}.
   At the special value $\phi = \pm \pi/6$, $\epsilon_{\uparrow} (K_i) = \epsilon_{\downarrow} (K'_i) =0$.
  Expanding near these points we obtain the cubic dispersion, typical for a higher-order VH singularity~\cite{Shtyk2017,*Classen2020PRB,*Classen2024}:
  \beq
  E_{\uparrow} ({\bf K}_i + {\bf k}) = E_{\downarrow} ({\bf K}'_i -{\bf k}) = \epsilon_k =
   \frac{|t|}{4} |k|^3 \cos (3 \psi_k),
 \label{ee_3}
 \eeq
 where $k_x = k \cos{\phi_k}$ and $k_y = k \sin{\phi_k}$.

 Within Hartree-Fock, the natural ordering tendency, both supported by a repulsive Hubbard $U$,  is towards a particle-hole order with either $Q=0$ or ${\bf Q}= 2{\bf K}_i$,  In our case, spin projection is locked,
  so fermions can be effectively treated as spinless. In this situation, a $Q=0$ particle-hole order  is valley polarization, while  $2{\bf K}_i$ order is an $XY$ antiferromagnetic order (XY AFM) because the z-component of spins at $K_i$ and $K{'}_i = -K_i$ are opposite.  Hartree-Fock calculations show~ \cite{MacDonald_TMD2018,DasSarma_TMD2020, Millis_TMD2021, Devakul:2021kn,Scheurer2024,*Scheurer2024_1,Guerci2024} that  valley polarization is the leading instability near $\phi = 0$, and XY AFM order wins  near  $\phi = \pi/6$. In the latter case, the instability develops  already  at arbitrary weak interaction because of the VH singularity.   The validity of Hartree-Fock calculations has, however, been questioned recently, particularly
   in the weak coupling regime near a VH point~\cite{Isobe2019,Classen,Schmalian,Shtyk2017,*Classen2020PRB,*Classen2024} because
   the VH-induced  increase of the interaction of the particle-hole channel
   can potentially be overcompensated by even stronger
   reduction due to  singular contributions from the particle-particle channel.

   \section{Superconducting Instability at weak coupling}
\label{Sec2}
   We begin by analyzing
    the possibility of a superconducting instability at weak coupling by computing the pairing vertex to second order in the 4-fermion interactions.
  In the next section we extend the calculations to stronger coupling and consider superconductivity in the parameter range
 where  the system is about to develop an XY AFM  order.

\begin{figure}
 \includegraphics[width= 0.5\columnwidth]{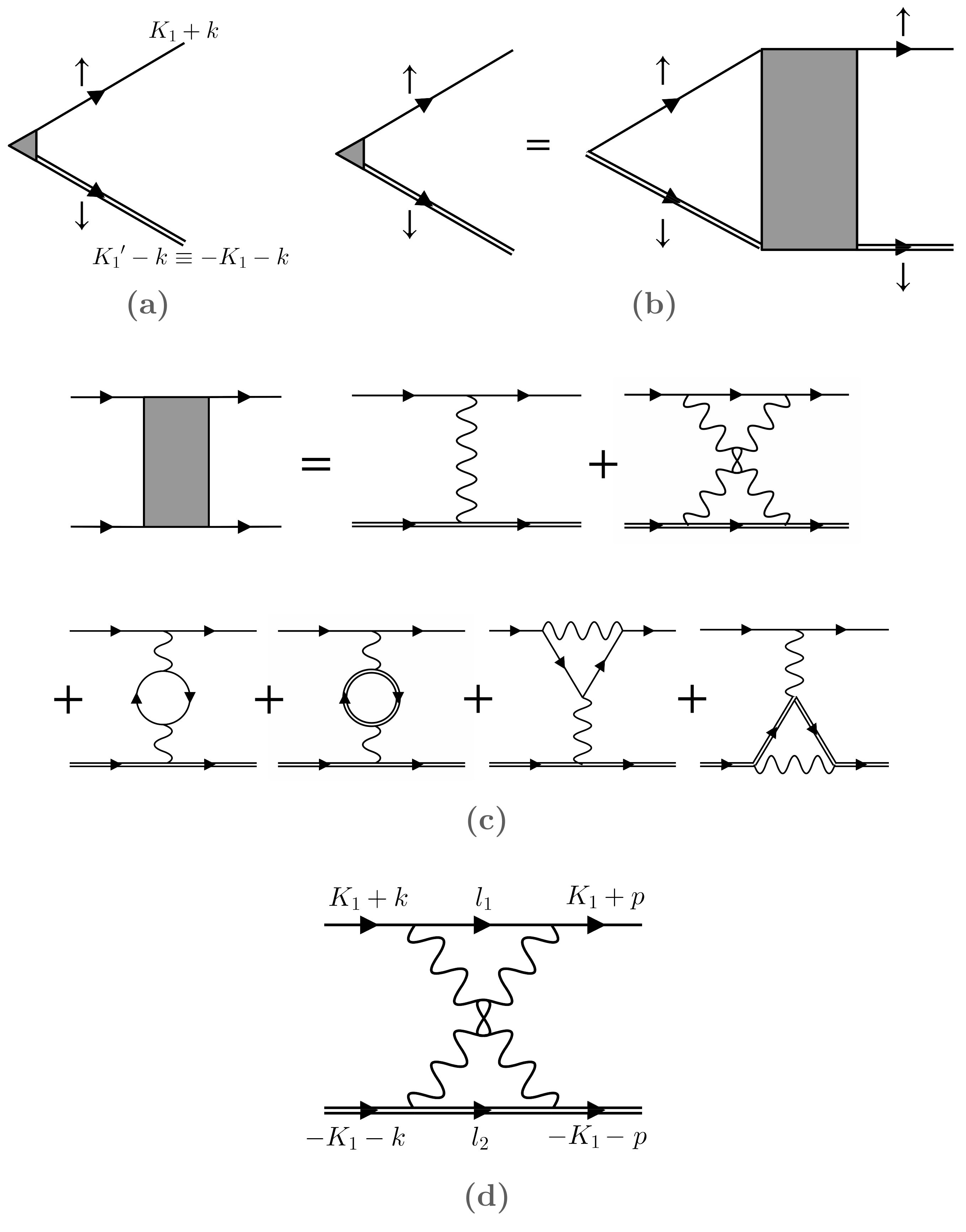}
\caption{Diagrammatic representation of (a) the gap function, (b) the linearized gap equation, (c) the irreducible vertex $\Gamma$ to second order in the interaction $U$ (depicted by wavy line) , and (d) the second-order crossed diagram. The two internal momenta ${\bf l}_1$ and ${\bf l}_2$ satisfy ${\bf l}_1 -{\bf l}_2 = 2{\bf K}_i + {\bf k} + {\bf p}$. }
 \label{fig:2}
\end{figure}

For definiteness, we consider  doping and displacement fields near the values
     where the van Hove singularity is at the Fermi level (the case
      $\phi = \pi/6$  in Eq. (\ref{bs})).
       If superconductivity exists there below a finite $T_c$, one expects it to continue at nearby values of $\phi$. We consider superconductivity with zero total momentum of a pair. Because $-{\bf K}_i = {\bf K}^{'}_i$, the gap function is then necessary  an $(\uparrow \downarrow)$   pair of an  $\uparrow$ fermion near $K_i$ and $\downarrow$ fermion near $K^{'}_i$ (Fig. \ref{fig:2}a).
  We show that
   the system develops   inversion-odd, two-component $(\uparrow \downarrow)$
   superconducting order ($E^{-}$ representation in notations of ~\cite{Scheurer2024,*Scheurer2024_1,Guerci2024})  by the Kohn-Luttinger (KL)  mechanism.
   \footnote{see ~\cite{KL,Raghu2010,kagan,*Chubukov93,Guinea2024,Schmalian} and references therein on KL scenario.}

We follow
~\cite{MacDonald_TMD2018},
and consider a
phenomenological
spin-preserving  repulsive  interaction $U$ between the density of low-energy $\uparrow$ fermions near ${\bf K}_i$ and the density of $\downarrow$ fermions near ${\bf K}{'}_i$. The important part of this interaction for the present purpose is near zero momentum transfer in the particle-hole channel as both the incoming and outgoing fermions  are either near ${\bf K}_i$ or near ${\bf K}^{'}_i$.  We neglect spin-preserving and spin-flip interactions with momentum transfer near $2{\bf K}_i$. For a  fully microscopic description one would need to consider gate-screened Coulomb interaction
   dressed by coherence factors associated with the projection onto the lowest-energy (topmost) bands, as has been recently done in ~\cite{Scheurer2024,*Scheurer2024_1,Guerci2024}. The coherence factors carry information about band topology.  Our rationale to restrict to a momentum independent interaction $U$ is that in our case the pairing comes exclusively from fermions
    in the immediate vicinity of $K$ and $K'$ points. For these fermions,  the coherence factors reduce to constants,
 which we incorporate into $U$.

  The linearized gap equation
  to second order in $U$ is shown diagrammatically in Fig.\ref{fig:2}b.  In analytical form,
 \beq
 \Delta_{{\bf k}} = -\int \frac{d^2p}{8\pi^2} \frac{\tanh{\frac{\epsilon_k}{2T_c}}}{\epsilon_k}  ~  \Gamma \left({\bf K}_i + {\bf k}, {\bf K}_i - {\bf k}, {\bf K'}_i + {\bf p}, {\bf K'}_i - {\bf p}, \right)~\Delta_{\bf p},
\label{ee_1}
\eeq
where $\epsilon_k$ is given by (\ref{ee_3}). The vertex $\Gamma$ with momentum indices as specified is a fully dressed  antisymmetrized irreducible pairing interaction.  One can easily make sure that in the absence of an interaction with momentum transfer $2{\bf K}_i$,  $\Gamma$ has no component with external fermions interchanged.  Its absence  implies that the dressed gap function retains an $\uparrow \downarrow$ structure, i.e.,  it is a mixture of spin-singlet and the $m_S =0$ part of the spin-triplet.  The diagrams for $\Gamma$ up to second order in the repulsive Hubbard $U$ are shown in  Fig. \ref{fig:2} c.  At the first order, $\Gamma$ is just $U$, at the second order in $U$, $\Gamma$  acquires a dependence on the momenta of the fermions.  Their angular dependence is present in all second-order diagrams, but because the low-energy excitations have a single spin component, the bubble and the vertex correction diagrams (i.e. Fig. (\ref{fig:2}) -c, second line) cancel  leaving only the "crossed" diagram (Fig. \ref{fig:2}d), which contains the polarizability diagram $\Pi_{ph} ({\bf Q})$ with ${\bf Q} \approx 2 {\bf K}_i$. In explicit form,
   $ \Gamma = U + U^2 ~ \Pi_{ph} (2{\bf K}_i +{\bf k} + {\bf p} )$ up to umklapp processes which give rise to a
     multiplication factor $3$,
     which we include into extending the angular integration near $K$ and $K'$ to the full circle.
         \begin{figure}
 \includegraphics[width= 0.5\columnwidth]{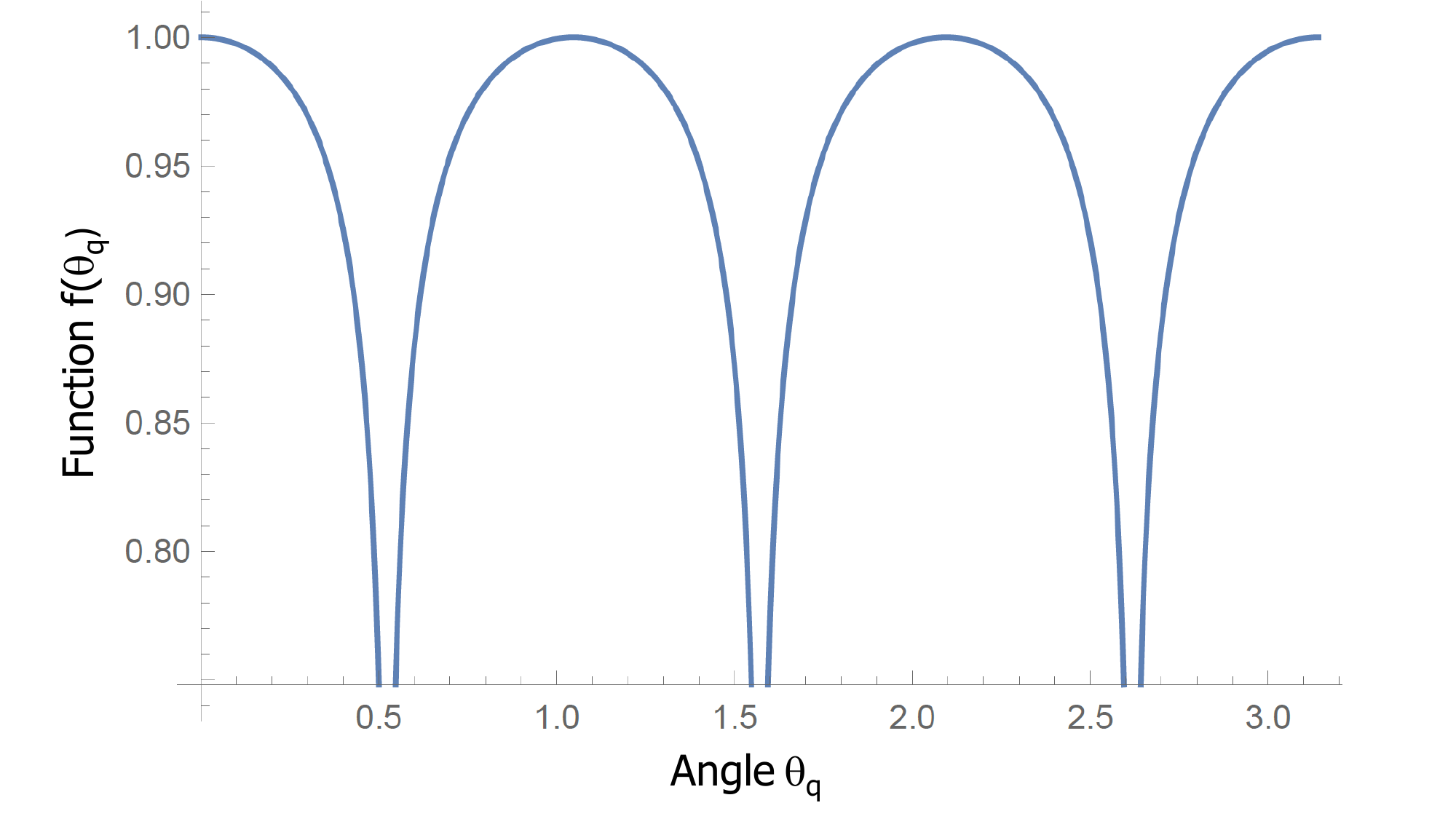}
\caption{The function $f(\theta_q)$ from Eq. (\ref{ee_16}).  The function is numerically quite close to one except for $\theta_q$ near $ (2n+1) \pi/6$, where $\cos{3\theta_q}$ vanishes. We argue in the text that superconductivity come from generic $\theta_k$ and is not confined to these special angles. We approximate $f(\theta_q)$ by one in the calculation of $T_c$.}
 \label{fig:function}
\end{figure}

        The $U$ term in $\Gamma$ is the repulsive interaction in the $s-$wave channel. The $U^2$  term is also repulsive, but is angular dependent and hence has non-zero components in finite angular momentum  channels.  If at least one  of these components is negative (attractive), the system is unstable towards superconductivity.   To check for this possibility we need to know the static particle-hole polarization bubble $\Pi_{ph}
        (2{\bf K}_i  + {\bf k} + {\bf p})$.  A simple analysis shows that $\Pi_{ph} (2{\bf K}_i)$  diverges.
          Evaluating it  for a finite
          ${\bf q} = {\bf k} + {\bf p} = (q \cos{\theta_q}, q \sin{\theta_q})$,
           we find that it scales as $1/q$ and continues to diverge for $\cos{3\theta_q} =0$.
           The generic expression for $\Pi ((2{\bf K}_i  + {\bf q})$ is
           \beq
          \Pi ((2{\bf K}_i  + {\bf q})= \frac{2}{3|t|} \frac{1}{|\cos{3\theta_q}|^{1/3}} f(\theta_q)
          \label{ee_16}
          \eeq
         where $f(n \pi/3) =1$ ($n$ is integer).  We plot $f(\theta_q)$ in Fig. \ref{fig:function}.  We see that it is  close to $1$ for most $\theta$ except tiny regions around $\theta_q = (2n+1) \pi/6$, where $\cos{3\theta_q}$ vanishes.  For these $\theta$, $\Pi ((2{\bf K}_i  + {\bf q})$ diverges as $|\log{|\cos{3\theta_q}|}|$. Still, even for such $\theta_q$,  $f(\theta_q)$ does not deviate strongly from one because numerically  $|\log{|\cos{3\theta_q}|}|$ and $1/|\cos{3\theta_q}|^{1/3}$ behave quite similarly except for a truly small range near $\theta_q = (2n+1) \pi/6$.   We will see that superconducting instability originates from $\cos{3\theta_q} = O(1)$. For these $\theta_q$, $f(\theta_q)$ can be well approximated by $1$, i.e., $ \Pi ((2{\bf K}_i  + {\bf q})= 2/(3|t||\cos{3\theta_q}|^{1/3})$. Substituting
          ${\bf q} = {\bf k} + {\bf p}$, we obtain
         \begin{equation}
    \Pi (2 {\bf K}_i + {\bf k}+ {\bf p}) = \frac{2}{3|t|} \frac{1}{|k^3 \cos{3\theta_k} + p^3 \cos{3\theta_p} + 3kp^2 \cos{(2\theta_p+\theta_k)} + 3pk^2 \cos{(2\theta_k+\theta_p)}|^{1/3}}
   \label{ee_9}
    \end{equation}

     A straightforward analysis of the gap equation shows that it decouples into separate equations for $\Delta_{{\bf k}}$ belonging to orthogonal irreducible representations. On a hexagonal lattice,
       these are 4 one-dimensional and 2 two-dimensional irreducible representations.  In the notations used for the case of no displacement field ($\phi =0$ in Eq. (\ref{bs}) the  one-dimensional irreducible representations  are labeled as $A_{1g}, A_{2g}, A_{1u}$, and $A_{2u}$ and the two-dimensional representations as  $E_g$ and $E_u$.
       $A_{1g}$ representation is a lattice version of an $s-$wave, other representations are lattice versions of  non-$s-$wave gap functions.  The gap functions in $A_{1g}, A_{2g}$ and $E_g$ are inversion-even and the ones in
         $A_{1u}, A_{2u}$ and $E_u$ are inversion-odd.

       Spin/lattice symmetry properties of $\Delta_{{\bf k}}$ in each representation have been discussed in~\cite{Scheurer2024,*Scheurer2024_1,Guerci2024}. In the presence of a displacement field, these properties get modified, and the representations  have been labeled as $E^{-}$, $E^+$ and $A^{-}_1$  instead of $E_u$, $E_g$ and
       $A_{1u}$, respectively, and so on. We will be using this notation below. For our purpose of solving the gap equation to find $T_c$, the essential fact is that for a lattice system  the  number of the gap components
       within a given irreducible representation is infinite. These components are specified in terms of $\theta_k$
       defined via ${\bf k} = k (\cos{\theta_k}, \sin{\theta_k})$. The specification is the same with and without the displacement field and is $\Delta ({\bf k}) \propto \cos{(3 (2n+1) \theta )}$ in $A_{1u}$ ($A^{-}_1$), where $n=0,1,2...$,
       $\Delta ({\bf k}) \propto \sin{(3 (2n+1) \theta )}$ in $A_{2u}$,  $\Delta ({\bf k}) \propto \cos{(6n \theta)}$ in $A_{1g}$ ($A^{+}_1$), $\Delta ({\bf k}) \propto \sin{(6n\theta )}$ in $A_{2g}$, $\Delta ({\bf k}) \propto \cos{(6n \pm 2) \theta )}$ in $E_{g}$ ($E^{+}$), and $\Delta ({\bf k}) \propto \cos{(6n \pm 1) \theta )}$ in $E_{u}$ ($E^{-}$).

     The  linearized integral equation on $\Delta_{\bf k}$ is obtained by substituting $\Pi (2 {\bf K}_i + {\bf k}+ {\bf p})$ given by (\ref{ee_9}) into (\ref{ee_1}).   In analytical form
     \be
     &&\Delta ({\bf k}) = -\frac{U^2}{6\pi^2 |t|^2} \int p dp d \theta_p \frac{\tanh{\frac{p^3 \cos{3\theta_p}}{2T^*_c}}}{p^3 \cos{3\theta_p}} \nonumber \\
     &&\times \frac{\Delta({\bf p})}{|k^3 \cos{3\theta_k} + p^3 \cos{3\theta_p} + 3kp^2 \cos{(2\theta_p+\theta_k)} + 3pk^2 \cos{(2\theta_k+\theta_p)}|^{1/3}}
     \label{ee_10}
     \ee
     where $T^*_c = 4T_c/|t|$.
      Note the overall minus sign -- the interaction at order $U^2$ is still a repulsion.

       Eq. (\ref{ee_10}) can be decoupled into a set of independent integral equations by choosing  $\Delta_{\bf k}$ and $\Delta (\bf p)$ to be within a given irreducible representation. Each gives its own $T_c$.
       Still, because there is an infinite number of partial components of $\Delta (\bf k)$  within a given representation, they all get mixed up, and for a generic $k$ the prefactors for the mixing terms
       are of order one. This apparently makes an analytic  progress difficult.  We argue, however, that this is possible by noting that the presence
      of $\tanh$ with the argument given in the r.h.s. of (\ref{ee_10}) implies that the natural scale for $p$ is  $p_0 \sim  (T^*_c)^{1/3}$.
  Consider external $k$ to be smaller than $p_0$ and  expand the denominator of (\ref{ee_10}) to leading order in $ k/p$ as
  \be
  &&\frac{1}{|k^3 \cos{3\theta_k} + p^3 \cos{3\theta_p} + 3kp^2 \cos{(2\theta_p+\theta_k)} + 3pk^2 \cos{(2\theta_k+\theta_p)}|^{1/3}} \nonumber \\
   &&\approx \frac{1}{p |\cos{3\theta_p}|^{1/3}} \left(1 - \frac{k}{p} \frac{\cos{(2\theta_p +\theta_k)}}{\cos{3\theta_p}}\right)
  \label{ee_11}
  \ee
  The first term is irrelevant as it just adds  to a repulsion in $A^{+}_1$ channel.  The second term appears with the minus sign and compensates the overall minus sign in (\ref{ee_10}). The $k$ dependence of this term is in the form $k \cos{\theta_k}$ or $k \sin{\theta_k}$. Both belong to $E^{-}$.  Assuming that $\Delta ({\bf p})$ for $p \sim p_0$ can be approximated by the same harmonics, we find after a straightforward calculation the set of two identical equations for  $T^*_c$ in the form
    \beq
     1 = \frac{U^2}{6\pi^2 t^2} \int_0^\infty dp  \int_0^{2\pi} d\theta_p \frac{\tanh{\frac{p^3 \cos{3\theta_p}}{2T^*_c}}}{p^3 \cos{3\theta_p}} \frac{1}{|\cos{3\theta_p}|^{1/3}}
    \label{ee_2}
    \eeq
    Introducing $u = p |\cos{3\theta_p}|^{1/3}$ as a new variable, integrating over $u$ and $\theta_p$ and using $\int_0^\infty du \tanh(u^3)/u^3 =1.435$, $\int_0^{2\pi} d \theta_p/|\cos{3\theta_p}|^{2/3} = 12 \sqrt{\pi}
    \Gamma(7/6)/\Gamma(2/3) = 14.572$, we obtain from (\ref{ee_2}) $T^*_c = 0.104(U/|t|)^3$, i.e.,
    \beq
    T_c = 0.026 |t| \left(\frac{U}{|t|}\right)^3
    \label{ee_12}
      \eeq
      We emphasize that $T_c$ has a power-law rather than exponential dependence on $U$ due to being in the vicinity of the van Hove singularity.
       We also note that the
      characteristic scale $p_0 \sim (T_c/|t|)^{1/3}$ is
      of order $U/|t|$. At weak coupling, this $p_0$ is much smaller than $|\bf K|$, which justifies the restriction to the vicinity of ${\bf K}$ and ${\bf K}'$.

We next consider the gap equation in the opposite limit $k \gg p_0$.   Here, we expand in $p/k$ and obtain
     \be
  &&\frac{1}{|k^3 \cos{3\theta_k} + p^3 \cos{3\theta_p} + 3kp^2 \cos{(2\theta_p+\theta_k)} + 3pk^2 \cos{(2\theta_k+\theta_p)}|^{1/3}} \nonumber \\
   &&\approx \frac{1}{k |\cos{3\theta_p}|^{1/3}} \left(1 - \frac{p}{k} \frac{\cos{(2\theta_k +\theta_p)}}{\cos{3\theta_k}}\right)
  \label{ee_14}
  \ee
  Taking the second term in the expansion to get an attractive interaction, we find that the gap function belongs to
  two-component, inversion-odd $E^{-}$ representation and its form is
  \beq
  \Delta ({\bf k}) \propto \frac{1}{k^2} \frac{ \cos{2\theta_k}}{\cos{3\theta_k}|\cos{3\theta_k}|^{1/3}} ~~{\text or}~~ \Delta ({\bf k}) \propto \frac{1}{k^2} \frac{ \sin{2\theta_k}}{\cos{3\theta_k}|\cos{3\theta_k}|^{1/3}}
  \eeq
  Substituting the same form for $\Delta (\bf p)$ into (\ref{ee_10}), like we did for small momenta, we find after simple algebra exactly the same equation for $T_c$ as in (\ref{ee_12}).

  The outcome of this study is that $E^{-}$ gap $\Delta ({\bf k})$ is a highly non-monotonic function of the magnitude of the momentum. At small $k$, it increases linearly with $k$, at large $k$ it decays as $1/k^2$.  The angular dependence also varies considerably. The crossover between the two regimes is at $k \sim p_0$, which, we remind, at weak coupling is  small compared to $|{\bf K}|$.

   A power-law dependence of $T_c$ on $U$ and a non-monotonic momentum dependence of the gap function has
        been also found in other studies of superconductivity near a higher-order VH point~\cite{Shtyk2017,*Classen2020PRB,*Classen2024,Schmalian}.  As we already stated, the gap function is a mixture  of spin-triplet/valley-triplet and spin-singlet/valley-singlet.

   The above calculation  of $T_c$  is approximate in two aspects.  First,  at both small $k \ll p_0$ and $k \gg p_0$, we  substituted $\Delta (\bf p)$ into the r.h.s. of (\ref{ee_10}) in the same form as  $\Delta ({\bf k})$. Meanwhile, the integral over $p$ is determined by $p \sim p_0$.  For such $p$, neither small $p$, nor large $p$ forms of $\Delta (\bf p)$ are accurate as other gap components from $E^{-}$ representation have comparable magnitudes (for small momentum expansion these components are  $k^5 \cos{5 \theta_k}$, $k^7\cos{7 \theta_k}$, etc).
  We conjecture that inclusion of these components changes the result for $T_c$, Eq. (\ref{ee_12}), by $O(1)$.
 Second,  we neglected the fermionic self-energy. It is singular at a
     a higher-order VH point and in principle may substantially affect  $T_c$~\cite{Schmalian}.
       To check whether this happens in our case, we evaluated the inverse quasiparticle residue $Z^{-1} (k, \omega) = d\Sigma (k, \omega)/d \omega$ to order $U^2$.  We found
       \beq
       Z^{-1} (k, \omega) = \left(\frac{U}{|t|}\right)^2 \frac{1}{k^2 |\cos{3 \psi_k}|^{2/3}}
        f\left(\frac{4\omega}{|t| k^3 |\cos{3 \psi_k}|}\right),
         \eeq
          where  $f(z = O(1)) = O(1)$.  Keeping this self-energy for the Green's functions in the pairing channel we obtained
    $T_c$ comparable to what we obtained neglecting the self-energy.  This is because  for  $|t| k^3 |\cos{3 \psi_k}| \sim \omega$, relevant for superconductivity, and $\omega \sim T_c$,  $Z(k, \omega) = O(1)$.
   We also remind that we consider a special case of VH singularity at ${\bf K}$ and ${\bf K}^{'}$, when pairing comes from fermions in the near vicinity of ${\bf K}$ and ${\bf K}^{'}$  For a generic case, pairing involves fermions in wider momentum ranges, for which  projection factors into the topmost valence band become relevant.

\begin{figure}
 \includegraphics[width= 0.5\columnwidth]{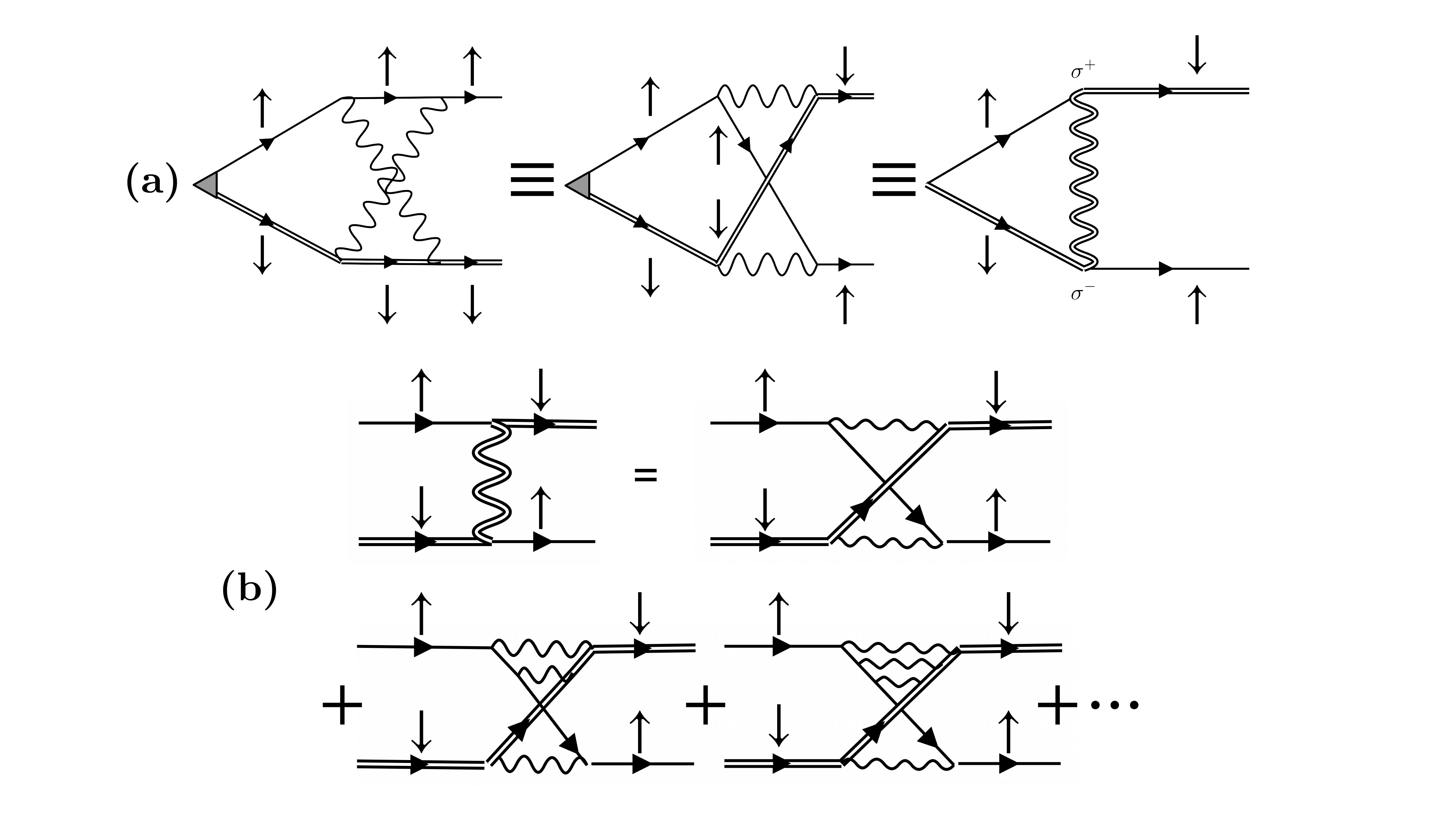}
\caption{Transformation from KL description to pairing mediated by XY AFM fluctuations.
(a) rearrangement of the pairing interaction at the order $U^2$ (the crossed diagram in Fig. \ref{fig:2}d) as the effective magnetically-mediated interaction with momentum transfer near $2{\bf K}_i$ and $\sigma_+$ and $\sigma_{-}$ in the side vertices. (b) Higher-order ladder renormalizations of the effective interaction. Earch ladder insertion contains interaction peaked at zero momentum transfer and the polarization bubble peaked at $2{\bf K}_i$. (c) Ladder series for the susceptibility of XY AFM order parameter. }
 \label{fig:4}
\end{figure}

The pairing gap  with  $E^{-}$  symmetry  has been obtained in \cite{Guerci2024} in the microscopic calculation for the model with gate-screened Coulomb interaction dressed by projection factors onto the low-energy bands. They also considered the case when the VH singularity is near the Fermi level.
Their  pairing mechanism is somewhat different from ours: they argued that the attraction comes from  RPA-type electronic screening of the Coulomb interaction. In  our model with the Hubbard-type interaction the screening contributions to order $U^2$ cancel out with the vertex correction diagrams (the ones in the second line in Fig. \ref{fig:2}c), while the attraction comes from the crossed diagram in the first line  in Fig. \ref{fig:2}c [We will show in the next Section that
 higher-order crossed diagrams  eventually lead to pairing mediate by XY AFM fluctuations].
   Despite this difference, the gap symmetry, which we obtained, fully agrees with the results by Guerci et al.  We view this equivalence as the evidence that the $E^{-}$ gap symmetry is robust and does not critically depends on the fine structure of the pairing interaction and on which processes yield an attractive pairing interaction.

    We also analyzed the gap equation in $E^{+}$ and $A_{1}^{-}$ channels. In both cases calculations require proper regularization of formally divergent terms.   In the $E^{+}$ channel ($\Delta ({\bf k}) \propto k^2 (\cos{2\theta_k}, \sin{2 \theta_k}))$   we found that the interaction remains repulsive, i.e., there is no pairing instability. In the
   $A_{1}^{-}$ channel ($\Delta ({\bf k}) \propto k^3 \cos{3\theta_k}$ at small $k$, we found an attraction, but $T_c$
    is smaller than the one in $E^{-}$ channel by $(2/3)^(3/2) = 0.544$.  As our calculations are only valid down to the largest $T_c$, the outcome of our calculation is pairing in $E^{-}$ channel.

Finally, the two-component structure of the $E^{-}$ gap function $\Delta ({\bf k}) = (\Delta_1 ({\bf k}), \Delta_2  ({\bf k}))$  raises the question  which combination the two gap components develops in a true superconducting state.
 To determine this, one has to go beyond $T_c$ and obtain Landau functional. One of the 4th order terms in this functional, of the form 
\be
W_4 (\Delta^*_1({\bf k }))^2 (\Delta({\bf k}))^2 + C.C., 
\ee
depends on the relative phase between the two components of $\Delta$.  When $W_4$ is positive, superconducting order is 
  chiral, with $\Delta ({\bf k}) = \Delta_1 ({\bf k}) \pm i \Delta_2 ({\bf k})$. Such an order also   breaks time-reversal symmetry.   A chiral state is expected on general grounds because it has a finite gap in its excitation spectra,  which normally leads to a larger condensation energy~Ref.~\cite{Nandkishore2012}.  Guerci et al solved the non-linear gap equation, derived the Landau  functional,  and found such a chiral order~\cite{Guerci2024}.

\subsection{RPA-type approach towards
 boson-mediated pairing}

We now extend the approach to stronger couplings $U \geq |t|$.
 At such couplings, the system may develop an instability towards a particle-hole order, or, at least, strong fluctuations of a particle-hole order parameter. For a two-valley system with low-energy fermions near ${\bf K}$ and ${\bf K}'$, candidates for a particle-hole order are valley polarization, intra-valley ferromagnetism, and
 XY AFM order with momentum ${\bf K} - {\bf K}'$ and  $\sigma^+$ ($\sigma^-$) structure of the order parameter.
 Hartree-Fock  calculations \cite{MacDonald_TMD2018,DasSarma_TMD2020, Millis_TMD2021, Devakul:2021kn,Scheurer2024,*Scheurer2024_1,Guerci2024} show that
  XY AFM order wins  near  $\phi = \pi/6$, where, we remind, the VH instability is at the Fermi level.
  For our model with equal intra-valley and inter-valley density-density interaction $U$ this holds because
   the dispersions $E_{\uparrow} ({\bf K}_i + {\bf k}) = - E_{\downarrow} ({\bf K}'_i +{\bf k})$, in which case a particle-hole bubble has the same structure as a particle-particle bubble and is enhanced.

Leaving aside the validity of Hartree-Fock approximation, we first show how one can obtain pairing mediated by XY AFM fluctuations peaked at $2{\bf K}_i$ even though in our model there is  no 4-fermion  interactions  with momentum transfer $2{\bf K}_i$. We discuss general aspects of pairing by $XY AFM$ fluctuations in the next section.

To relate pairing interaction  to XY magnetic fluctuations at $2{\bf K}_i$ we  note that
the pairing vertex in Fig. \ref{fig:2}d (the crossed diagram)  contains a particle-hole bubble with momentum near $2{\bf K}_i$.  Using this, we re-express the crossed diagram as the effective interaction between $\uparrow$ fermion near ${\bf K}_i$ and $\downarrow$ fermion near ${\bf K}^{'}_i$, as shown in  Fig.\ref{fig:4}a.
The pairing interaction $V$ is the same as before: $V = U^2 \Pi_{ph} (2{\bf K}_i  + {\bf k} + {\bf p})$, but
   because fermions along the interaction line now have  $\uparrow$ and $\downarrow$ spin projections, it  can be viewed as the magnetically-mediated interaction peaked at $2{\bf K}_i$,  with $\sigma^{+} =(\sigma_x + i \sigma_y)/2$ and $\sigma^{-} = (\sigma_x - i \sigma_y)/2$ in the side vertices.

   To order $U^2$, this is just a rearrangement. However, at larger $U \geq |t|$, higher-loop renormalizations become relevant. We follow earlier work~\cite{Dong2024} and add ladder series of particle-hole bubbles, as shown in Fig.\ref{fig:4}b.
   The interactions in the ladder series are still with momentum transfer near zero, but the bubbles are with momentum transfer near $2{\bf K}_i$.  The reasoning for the diagrammatic selection is that the same ladder series also account for the RPA-type enhancement of the susceptibility of the XY AFM order, Fig. \ref{fig:4}c.
   Comparing the two series, we find that  the ladder renormalization of the pairing interaction  $V$
     multipies it by the
    susceptibility of the XY $2{\bf K}_i-$order parameter.  The outcome is that near the onset of  XY AFM order, when the XY AFM susceptibility is large,  the pairing can be viewed as  mediated by soft XY magnetic  fluctuations peaked at a finite momentum $2{\bf K}_i$.  The selection of ladder diagrams cannot be rigorously justified at strong coupling, and the derivation of the effective interaction should be treated as an illustration. Yet, it provides a continuous transformation between KL-type superconductivity at weak coupling and magnetically-mediated pairing at strong coupling.  Because the evolution is continuous, the gap symmetry should remain the same  $E^-$, i.e., superconducting  gap function remains  two-component, inversion-odd, with $\uparrow, \downarrow$  spin structure, i.e. a mixture between spin triplet and spin singlet.

\section{Strong coupling: Pairing by fluctuations of \\ XY AFM order.}
\label{Sec3}

  Pairing by AFM fluctuations has been extensively discussed  following  the first calculations based on the  symmetry of the spatial and spin fluctuations \cite{MiyakeSV1986} and the RPA-typle calculations \cite{Scalapino1986} in connection with heavy-fermion superconductivity. They showed that  $SU(2)$-invariant AFM fluctuations promote  even parity-singlet superconductivity, specifically d-wave pairing in cubic or tetragonal crystals. In our case, this needs to be considered for $U(1)$ fluctuations relevant to the present problem.

In general, the coupling between fermions and fluctuations of the XY AFM order is described by
\beq
H_i= g~
\sum_i
M^{-} ({\bf K}_i - {\bf K}^{'}_i +{\bf k} +{\bf p})
c^+_\alpha ({\bf K}_i + {\bf k}) \sigma^{+}_{\alpha \beta} c_\beta ({\bf K}^{'}_i -{\bf p}) + h.c.
\label{fmcoupl}
\eeq
 where $M^{-}, M^{+} = M_x \mp i M_y$ are conjugate transverse magnetic  moments and $g$ is a function of $U$ and $t$.
Consider the symmetry of superconductivity promoted by (\ref{fmcoupl}).   The basic irreducible vertex $I(K_i ,K'_i , q, \omega)$ in the particle-hole channel is proportional to the fluctuation propagator $<M^+ M^->(2 {\bf K_i +q}, \omega)$ given below.
This provides an attractive interaction in the particle-hole channel  for ${\bf k}$ near $2 {\bf K_i}$ since $Re ~I < 0$ below the upper cut-off $\omega_c$ of the fluctuations. In the particle-particle channel the sign is switched.  For isotropic interactions in spin-space, the sign further depends on whether pairing in singlet where it provides, on taking the spin-trace, an attractive pairing interaction,  or triplet where the trace provides a repulsive interaction \cite{MiyakeSV1986}.  In our case, there is no such choice since the spin-states for large angle scattering have been fixed by the spin-orbit coupling. The pairing then is as already obtained above, a linear combination of the singlet and triplet states.

 To get an insight about pairing in this situation, let us
 consider a toy model with a circular Fermi surface
   and isotropic dispersion for fermions near each pocket.
    The modification of the procedure leading to Eq. (2) of Ref. \cite{MiyakeSV1986} specialized to the present case in which there is no spin-trace to consider gives the  pairing interaction $V_{\ell}$ in the $\ell$-th angular momentum channel. For  the integration variable $x = \sin \theta/2$, where $\theta$ is the angle between pairs of particles on the Fermi-surface,
\be
V_{\ell} = 2 g^2 \int_0^1 dx P_{\ell}(1-2x^2) I(2 k_Fx, \uparrow, \downarrow).
\ee
Clearly, if the fluctuations $I(2 k_Fx)$ peaks at $\theta = \pi$, only odd angular momentum provides attractive
interactions.   The leading candidates are two-component $\ell =1$ and $\ell =3$ gap functions ($\cos{\theta}/\sin{\theta}$ with two lines of zero and $\cos{3\theta}/\sin{3\theta}$ with six lines of zeroes.
The $\ell =1$  gap structures is a  prototype for the $E^{-}$ gap  and the $\ell =3$ gap structure is a prototype for $A_{1}^{-}$ gap.

To solve the gap equation to determine $T_c$, we need to know the
form of the dynamical susceptibility for XY AFM fluctuations coupled to fermions.  It has been argued by one of us (see, e.g., ~\cite{VarmaRMP2020})  that the
   full expression for the dynamical propagator of XY fluctuations must include topological contributions from  vortices and is different from a conventional diagrammatic expression with Landau damping.
    For this reason, we will use the form of the  correlator of $M_+$ and $M_-$, which was extracted as a function of  coordinates and the  imaginary time from
 quantum-Monte-carlo and renormalization group calculations \cite{ZhuChenV2015, ZhuHouV2016, Hou-CMV-RG}.
  The correlator as a function of  momentum and real frequency is obtained after Fourier transform and  analytic continuation.  At XY AFM criticality, it  is
\be
\label{corr1}
Im \langle M_+ M_-(0,0)\rangle (2{\bf K}_i + {\bf q}, ( \omega, T)
\approx  = \chi_0 \tanh\big(\frac{\omega}{2T}\big) \frac{1}{(2{\bf K}_i + {\bf q})^2},
 \ee
where  $\chi_0$ gives the amplitude of the fluctuations. Away from criticality, the $\omega/T$ dependence is cut-off by an inverse correlation length in time $\xi_{\tau}^{-1}$ and the singularity in momentum by an inverse correlation length in  space $\xi_r^{-1}$.  The spatial correlation length  is logarithm of the temporal correlation length so that the correlations are effectively local in space, especially for AFM fluctuations. The detailed forms are given in
 Eqs. (26-27) in \cite{ZhuHouV2016}.
 These properties appear to be essential for obtaining the experimentally observed transport and thermodynamic properties in the critical region. It should be noted that for the two metals in which neutron scattering results are available in the AFM quantum-critical regime, CeCu$_{5.9}$Au$_{0.1}$ (Ref. ~\cite{Schroeder2000}) and  BaFe$_{1.85}$Co$_{0.15}$As$_2$ (Ref.~\cite{Inosov2010}), analysis of the data \cite{CMVZhuSchroeder2016} reveals that the fluctuation spectra is consistent with the form in Eq. (\ref{corr1}).
The inversion of the angle-resolved photoemission spectra in the cuprate compound BISCCO \cite{Bok_ScienceADV}. also shows that fluctuations of this form lead to the anomalous normal state scattering rate as well as the d-wave superconductivity. In the present context, the linear in $H$ resistivity \cite{Varma_RH2022} recently found in WSe$_2$ beside the linear in $T$ resistivity \cite{Mak_supercond_2024} are especially noteworthy in this regard.  We will not dwell on the  application of the fluctuations (\ref{corr1}) to the normal state properties as they have been described  elsewhere \cite{Varma_IOPrev2016}.

For superconductivity, the factorization of the pairing interaction into momentum and frequency dependent parts implies that the momentum dependence determines the gap symmetry, while the frequency dependence determines the value of $T_c$.
 The momentum dependence of the pairing interaction does not critically depend on the inclusion of the topological excitations and is qualitatively the same as of the static XY susceptibility obtained by conventional diagrammatic means.  We use Refs. \cite{Scheurer2024,*Scheurer2024_1}, who computed the gap symmetry  using  a phenomenological model of pairing mediated by XY AFM fluctuations and diagrammatic expression for the static XY propagator with the coherence factors of projection into low-energy fermionic states.  They found $E^{-}$ gap structure, the same as at weak coupling, with $A^{-}$ channel also attractive, but weaker.

To determine $T_c$ and its dependence of the coupling, we the frequency dependent factor in the real part of the XY susceptibility (Kramers-Kronig transform of  Eq. (\ref{corr1})).
 The latter scales as $\log{\omega_c/|\omega_m}$ at frequencies below $\omega_c$, which is generally of order $|t|$~\cite{Varma_IOPrev2016}. The superconductivity singularity is then of the $log^2(\omega_c/T)$ form. The equation of the frequency-dependent gap is then
\beq
\Delta (\omega_m) = \lambda \int_0^{\omega_{c}} \frac{d \omega^{'}_m}{\omega^{'}_m} \log{\frac{\omega_c}{|\omega^{'}_m|}}
\eeq
 where the dimensionless $\lambda \propto (g/t)^2$ and it also includes the numerical factor coming from solving the gap equation in momentum space.   If  $\lambda$ is small, $T_c \propto \omega_c e^{-\pi/2\sqrt{\lambda}}$ (Ref. \cite{son}).  For $\lambda = O(1)$, $T_c$ is comparable to $|t|$.

 A caveat on the use of the form (\ref{corr1}) to the present problem should be mentioned. Eq. (\ref{corr1}) is derived for the fluctuations of an order (with power law correlations)
 on a bipartite lattice. But the magnetic order  in t-WSe$_2$ is expected to be of the form of  three sub-lattice order on a triangular lattice with spins-pointing at $2\pi/3$ with respect to each other. The problem of fluctuations for such an order has not even been solved classically,
  although excellent classical Monte-carlo calculations are available \cite{Shiba1984}. The eventual order can lead to two kinds of $2\pi/3$ order differing in chirality. Therefore both fluctuations of the Kosterlitz-Thouless kind for the xy model and that of an Ising transition are expected. The former however dominate in the results obtained~ \cite{Shiba1984}.
  No results, analytical or from quantum-Monte-carlo calculations are available
  for the quantum XY problem.   We have assumed that in the quantum
   case
   the fluctuations of the xy  transition dominate, guided by the fact that they explain the observed normal state anomalies.
 For this as well as other similar problems firm results for the xy model on a triangular lattice are desirable.

\section{Concluding Remarks.}
\label{Sec4}
We have analyzed superconductivity in twisted transition metal chalcogenide WSe$_2$ (t-WSe$_2$), where
 spin-orbit scattering locks the spins of low-energy excitations near Dirac ${\bf K}$ and ${\bf K}'$ points.
 We show that a nominally repulsive 4-fermion interaction gives rise to an attraction in a two-component $E^{-}$ channel. The gap function is inversion-odd and its spin structure is $\uparrow \downarrow$, i.e., it is a mixture between spin singlet and spin triplet.  At weak coupling, $E^-$  superconductivity emerges via the Kohn-Luttinger mechanism. We computed the corresponding $T_c$ for the case when van Hove singularity is near the Fermi level and found that it has a power-law dependence on the interaction.   At strong coupling, the pairing is mediated by XY magnetic  fluctuations peaked at momenta ${\bf K}_i-{\bf K}' _i= 2 {\bf K}_i$. From pure symmetry considerations one finds the same state because the symmetry depends primarily on the momentum dependence of the interactions and the electronic structure.
 We then used the form of the propagator that displays $\omega/T$ scaling, suggested and derived earlier for the xy model coupled to fermion; for a review see \cite{Varma_IOPrev2016}. Due to the  $\omega/T$ scaling  of the fluctuations, one finds a $| log^2 \omega_c/\omega$ singularity for pairing.  When the cut-off $\omega_c$ of the fluctuations  and the bandwidth are comparable,  $T_c$ scales with the bandwidth. The same fluctuations scattered by fermions give the observed normal state anomalies unlike fluctuations calculated in RPA.

 {\it Acknowledgements.}~~ CMV acknowledges important discussions with Liang Fu on the applicability of xy- criticality in twisted WSe$_2$ and to him and to Trithep Devakul  and Andrew Millis on discussions on the electronic structure of this compound and details of their calculations on it. Discussions of the experimental data with
Augusto Ghiotto, Kin Fai Mak and A. Pashupathy were very helpful. AVC acknowledges with thanks discussions with E. Berg, L. Classen, M. Kokkinis, D. Mayrhofer, M. Scheurer and J. Schmalian.
 The work of AVC was supported by the U.S. Department of Energy, Office of Science, Basic Energy Sciences,
under Award No. DE-SC0014402. Part of this work was done at the Aspen Center for Physics which is partially supported by the NSF through grant PHY-1607611. Part of this work was done at the Aspen Center for Physics, which is supported in part by PHY-2210452.

\bibliography{TMD.bib}

\begin{thebibliography}{57}%
\makeatletter
\providecommand \@ifxundefined [1]{%
 \@ifx{#1\undefined}
}%
\providecommand \@ifnum [1]{%
 \ifnum #1\expandafter \@firstoftwo
 \else \expandafter \@secondoftwo
 \fi
}%
\providecommand \@ifx [1]{%
 \ifx #1\expandafter \@firstoftwo
 \else \expandafter \@secondoftwo
 \fi
}%
\providecommand \natexlab [1]{#1}%
\providecommand \enquote  [1]{``#1''}%
\providecommand \bibnamefont  [1]{#1}%
\providecommand \bibfnamefont [1]{#1}%
\providecommand \citenamefont [1]{#1}%
\providecommand \href@noop [0]{\@secondoftwo}%
\providecommand \href [0]{\begingroup \@sanitize@url \@href}%
\providecommand \@href[1]{\@@startlink{#1}\@@href}%
\providecommand \@@href[1]{\endgroup#1\@@endlink}%
\providecommand \@sanitize@url [0]{\catcode `\\12\catcode `\$12\catcode
  `\&12\catcode `\#12\catcode `\^12\catcode `\_12\catcode `\%12\relax}%
\providecommand \@@startlink[1]{}%
\providecommand \@@endlink[0]{}%
\providecommand \url  [0]{\begingroup\@sanitize@url \@url }%
\providecommand \@url [1]{\endgroup\@href {#1}{\urlprefix }}%
\providecommand \urlprefix  [0]{URL }%
\providecommand \Eprint [0]{\href }%
\providecommand \doibase [0]{https://doi.org/}%
\providecommand \selectlanguage [0]{\@gobble}%
\providecommand \bibinfo  [0]{\@secondoftwo}%
\providecommand \bibfield  [0]{\@secondoftwo}%
\providecommand \translation [1]{[#1]}%
\providecommand \BibitemOpen [0]{}%
\providecommand \bibitemStop [0]{}%
\providecommand \bibitemNoStop [0]{.\EOS\space}%
\providecommand \EOS [0]{\spacefactor3000\relax}%
\providecommand \BibitemShut  [1]{\csname bibitem#1\endcsname}%
\let\auto@bib@innerbib\@empty
\bibitem [{\citenamefont {Cao}\ \emph {et~al.}(2018{\natexlab{a}})\citenamefont
  {Cao}, \citenamefont {Fatemi}, \citenamefont {Demir}, \citenamefont {Fang},
  \citenamefont {Tomarken}, \citenamefont {Luo}, \citenamefont
  {Sanchez-Yamagishi}, \citenamefont {Watanabe}, \citenamefont {Taniguchi},
  \citenamefont {Kaxiras}, \citenamefont {Ashoori},\ and\ \citenamefont
  {Jarillo-Herrero}}]{Cao:2018xq}%
  \BibitemOpen
  \bibfield  {author} {\bibinfo {author} {\bibfnamefont {Y.}~\bibnamefont
  {Cao}}, \bibinfo {author} {\bibfnamefont {V.}~\bibnamefont {Fatemi}},
  \bibinfo {author} {\bibfnamefont {A.}~\bibnamefont {Demir}}, \bibinfo
  {author} {\bibfnamefont {S.}~\bibnamefont {Fang}}, \bibinfo {author}
  {\bibfnamefont {S.~L.}\ \bibnamefont {Tomarken}}, \bibinfo {author}
  {\bibfnamefont {J.~Y.}\ \bibnamefont {Luo}}, \bibinfo {author} {\bibfnamefont
  {J.~D.}\ \bibnamefont {Sanchez-Yamagishi}}, \bibinfo {author} {\bibfnamefont
  {K.}~\bibnamefont {Watanabe}}, \bibinfo {author} {\bibfnamefont
  {T.}~\bibnamefont {Taniguchi}}, \bibinfo {author} {\bibfnamefont
  {E.}~\bibnamefont {Kaxiras}}, \bibinfo {author} {\bibfnamefont {R.~C.}\
  \bibnamefont {Ashoori}},\ and\ \bibinfo {author} {\bibfnamefont
  {P.}~\bibnamefont {Jarillo-Herrero}},\ }\bibfield  {title} {\bibinfo {title}
  {Correlated insulator behaviour at half-filling in magic-angle graphene
  superlattices},\ }\href {https://doi.org/10.1038/nature26154} {\bibfield
  {journal} {\bibinfo  {journal} {Nature}\ }\textbf {\bibinfo {volume} {556}},\
  \bibinfo {pages} {80} (\bibinfo {year} {2018}{\natexlab{a}})}\BibitemShut
  {NoStop}%
\bibitem [{\citenamefont {Cao}\ \emph {et~al.}(2018{\natexlab{b}})\citenamefont
  {Cao}, \citenamefont {Fatemi}, \citenamefont {Fang}, \citenamefont
  {Watanabe}, \citenamefont {Taniguchi}, \citenamefont {Kaxiras},\ and\
  \citenamefont {Jarillo-Herrero}}]{Cao:2018lk}%
  \BibitemOpen
  \bibfield  {author} {\bibinfo {author} {\bibfnamefont {Y.}~\bibnamefont
  {Cao}}, \bibinfo {author} {\bibfnamefont {V.}~\bibnamefont {Fatemi}},
  \bibinfo {author} {\bibfnamefont {S.}~\bibnamefont {Fang}}, \bibinfo {author}
  {\bibfnamefont {K.}~\bibnamefont {Watanabe}}, \bibinfo {author}
  {\bibfnamefont {T.}~\bibnamefont {Taniguchi}}, \bibinfo {author}
  {\bibfnamefont {E.}~\bibnamefont {Kaxiras}},\ and\ \bibinfo {author}
  {\bibfnamefont {P.}~\bibnamefont {Jarillo-Herrero}},\ }\bibfield  {title}
  {\bibinfo {title} {Unconventional superconductivity in magic-angle graphene
  superlattices},\ }\href {https://doi.org/10.1038/nature26160} {\bibfield
  {journal} {\bibinfo  {journal} {Nature}\ }\textbf {\bibinfo {volume} {556}},\
  \bibinfo {pages} {43} (\bibinfo {year} {2018}{\natexlab{b}})}\BibitemShut
  {NoStop}%
\bibitem [{\citenamefont {Andrei}\ \emph {et~al.}(2021)\citenamefont {Andrei},
  \citenamefont {Efetov}, \citenamefont {Jarillo-Herrero}, \citenamefont
  {MacDonald}, \citenamefont {Mak}, \citenamefont {Senthil}, \citenamefont
  {Tutuc}, \citenamefont {Yazdani},\ and\ \citenamefont
  {Young}}]{EAndreiREV2021}%
  \BibitemOpen
  \bibfield  {author} {\bibinfo {author} {\bibfnamefont {E.}~\bibnamefont
  {Andrei}}, \bibinfo {author} {\bibfnamefont {D.}~\bibnamefont {Efetov}},
  \bibinfo {author} {\bibfnamefont {P.}~\bibnamefont {Jarillo-Herrero}},
  \bibinfo {author} {\bibfnamefont {A.}~\bibnamefont {MacDonald}}, \bibinfo
  {author} {\bibfnamefont {K.}~\bibnamefont {Mak}}, \bibinfo {author}
  {\bibfnamefont {T.}~\bibnamefont {Senthil}}, \bibinfo {author} {\bibfnamefont
  {E.}~\bibnamefont {Tutuc}}, \bibinfo {author} {\bibfnamefont
  {A.}~\bibnamefont {Yazdani}},\ and\ \bibinfo {author} {\bibfnamefont
  {A.}~\bibnamefont {Young}},\ }\bibfield  {title} {\bibinfo {title} {The
  marvels of moir{\'e} materials},\ }\href
  {https://doi.org/10.1038/s41578-021-00284-1} {\bibfield  {journal} {\bibinfo
  {journal} {Nature Reviews Materials}\ }\textbf {\bibinfo {volume} {6}},\
  \bibinfo {pages} {201} (\bibinfo {year} {2021})}\BibitemShut {NoStop}%
\bibitem [{\citenamefont {Ghiotto}\ \emph {et~al.}(2021)\citenamefont
  {Ghiotto}, \citenamefont {Shih}, \citenamefont {Pereira}, \citenamefont
  {Rhodes}, \citenamefont {Kim}, \citenamefont {Zang}, \citenamefont {Millis},
  \citenamefont {Watanabe}, \citenamefont {Taniguchi}, \citenamefont {Hone},
  \citenamefont {Wang}, \citenamefont {Dean},\ and\ \citenamefont
  {Pasupathy}}]{Pasupathy2021}%
  \BibitemOpen
  \bibfield  {author} {\bibinfo {author} {\bibfnamefont {A.}~\bibnamefont
  {Ghiotto}}, \bibinfo {author} {\bibfnamefont {E.-M.}\ \bibnamefont {Shih}},
  \bibinfo {author} {\bibfnamefont {G.~S. S.~G.}\ \bibnamefont {Pereira}},
  \bibinfo {author} {\bibfnamefont {D.~A.}\ \bibnamefont {Rhodes}}, \bibinfo
  {author} {\bibfnamefont {B.}~\bibnamefont {Kim}}, \bibinfo {author}
  {\bibfnamefont {J.}~\bibnamefont {Zang}}, \bibinfo {author} {\bibfnamefont
  {A.~J.}\ \bibnamefont {Millis}}, \bibinfo {author} {\bibfnamefont
  {K.}~\bibnamefont {Watanabe}}, \bibinfo {author} {\bibfnamefont
  {T.}~\bibnamefont {Taniguchi}}, \bibinfo {author} {\bibfnamefont {J.~C.}\
  \bibnamefont {Hone}}, \bibinfo {author} {\bibfnamefont {L.}~\bibnamefont
  {Wang}}, \bibinfo {author} {\bibfnamefont {C.~R.}\ \bibnamefont {Dean}},\
  and\ \bibinfo {author} {\bibfnamefont {A.~N.}\ \bibnamefont {Pasupathy}},\
  }\bibfield  {title} {\bibinfo {title} {Quantum criticality in twisted
  transition metal dichalcogenides},\ }\href
  {https://doi.org/10.1038/s41586-021-03815-6} {\bibfield  {journal} {\bibinfo
  {journal} {Nature}\ }\textbf {\bibinfo {volume} {597}},\ \bibinfo {pages}
  {345} (\bibinfo {year} {2021})}\BibitemShut {NoStop}%
\bibitem [{\citenamefont {Xu}\ \emph {et~al.}(2022)\citenamefont {Xu},
  \citenamefont {Kang}, \citenamefont {Watanabe}, \citenamefont {Taniguchi},
  \citenamefont {Mak},\ and\ \citenamefont {Shan}}]{Mak_WSE2_2022}%
  \BibitemOpen
  \bibfield  {author} {\bibinfo {author} {\bibfnamefont {Y.}~\bibnamefont
  {Xu}}, \bibinfo {author} {\bibfnamefont {K.}~\bibnamefont {Kang}}, \bibinfo
  {author} {\bibfnamefont {K.}~\bibnamefont {Watanabe}}, \bibinfo {author}
  {\bibfnamefont {T.}~\bibnamefont {Taniguchi}}, \bibinfo {author}
  {\bibfnamefont {K.~F.}\ \bibnamefont {Mak}},\ and\ \bibinfo {author}
  {\bibfnamefont {J.}~\bibnamefont {Shan}},\ }\bibfield  {title} {\bibinfo
  {title} {A tunable bilayer hubbard model in twisted wse2},\ }\href
  {https://doi.org/10.1038/s41565-022-01180-7} {\bibfield  {journal} {\bibinfo
  {journal} {Nature Nanotechnology}\ }\textbf {\bibinfo {volume} {17}},\
  \bibinfo {pages} {934} (\bibinfo {year} {2022})}\BibitemShut {NoStop}%
\bibitem [{\citenamefont {Lu}\ \emph {et~al.}(2019)\citenamefont {Lu},
  \citenamefont {Stepanov}, \citenamefont {Yang}, \citenamefont {Xie},
  \citenamefont {Aamir}, \citenamefont {Das}, \citenamefont {Urgell},
  \citenamefont {Watanabe}, \citenamefont {Taniguchi}, \citenamefont {Zhang},
  \citenamefont {Bachtold}, \citenamefont {MacDonald},\ and\ \citenamefont
  {Efetov}}]{Efetov2019}%
  \BibitemOpen
  \bibfield  {author} {\bibinfo {author} {\bibfnamefont {X.}~\bibnamefont
  {Lu}}, \bibinfo {author} {\bibfnamefont {P.}~\bibnamefont {Stepanov}},
  \bibinfo {author} {\bibfnamefont {W.}~\bibnamefont {Yang}}, \bibinfo {author}
  {\bibfnamefont {M.}~\bibnamefont {Xie}}, \bibinfo {author} {\bibfnamefont
  {M.~A.}\ \bibnamefont {Aamir}}, \bibinfo {author} {\bibfnamefont
  {I.}~\bibnamefont {Das}}, \bibinfo {author} {\bibfnamefont {C.}~\bibnamefont
  {Urgell}}, \bibinfo {author} {\bibfnamefont {K.}~\bibnamefont {Watanabe}},
  \bibinfo {author} {\bibfnamefont {T.}~\bibnamefont {Taniguchi}}, \bibinfo
  {author} {\bibfnamefont {G.}~\bibnamefont {Zhang}}, \bibinfo {author}
  {\bibfnamefont {A.}~\bibnamefont {Bachtold}}, \bibinfo {author}
  {\bibfnamefont {A.~H.}\ \bibnamefont {MacDonald}},\ and\ \bibinfo {author}
  {\bibfnamefont {D.~K.}\ \bibnamefont {Efetov}},\ }\bibfield  {title}
  {\bibinfo {title} {Superconductors, orbital magnets and correlated states in
  magic-angle bilayer graphene},\ }\href
  {https://doi.org/10.1038/s41586-019-1695-0} {\bibfield  {journal} {\bibinfo
  {journal} {Nature}\ }\textbf {\bibinfo {volume} {574}},\ \bibinfo {pages}
  {653} (\bibinfo {year} {2019})}\BibitemShut {NoStop}%
\bibitem [{\citenamefont {Xia}\ \emph {et~al.}(2024)\citenamefont {Xia},
  \citenamefont {Han}, \citenamefont {Watanabe}, \citenamefont {Taniguchi},
  \citenamefont {Shan},\ and\ \citenamefont {Mak}}]{Mak_supercond_2024}%
  \BibitemOpen
  \bibfield  {author} {\bibinfo {author} {\bibfnamefont {Y.}~\bibnamefont
  {Xia}}, \bibinfo {author} {\bibfnamefont {Z.}~\bibnamefont {Han}}, \bibinfo
  {author} {\bibfnamefont {K.}~\bibnamefont {Watanabe}}, \bibinfo {author}
  {\bibfnamefont {T.}~\bibnamefont {Taniguchi}}, \bibinfo {author}
  {\bibfnamefont {J.}~\bibnamefont {Shan}},\ and\ \bibinfo {author}
  {\bibfnamefont {K.~F.}\ \bibnamefont {Mak}},\ }\href
  {https://arxiv.org/abs/2405.14784} {\bibinfo {title} {Unconventional
  superconductivity in twisted bilayer wse2}} (\bibinfo {year} {2024}),\
  \Eprint {https://arxiv.org/abs/2405.14784} {arXiv:2405.14784
  [cond-mat.mes-hall]} \BibitemShut {NoStop}%
\bibitem [{\citenamefont {Guo}\ \emph {et~al.}(2024)\citenamefont {Guo},
  \citenamefont {Pack}, \citenamefont {Swann}, \citenamefont {Holtzman},
  \citenamefont {Cothrine}, \citenamefont {Watanabe}, \citenamefont
  {Taniguchi}, \citenamefont {Mandrus}, \citenamefont {Barmak}, \citenamefont
  {Hone}, \citenamefont {Millis}, \citenamefont {Pasupathy},\ and\
  \citenamefont {Dean}}]{Pasupathy2024superconductivity}%
  \BibitemOpen
  \bibfield  {author} {\bibinfo {author} {\bibfnamefont {Y.}~\bibnamefont
  {Guo}}, \bibinfo {author} {\bibfnamefont {J.}~\bibnamefont {Pack}}, \bibinfo
  {author} {\bibfnamefont {J.}~\bibnamefont {Swann}}, \bibinfo {author}
  {\bibfnamefont {L.}~\bibnamefont {Holtzman}}, \bibinfo {author}
  {\bibfnamefont {M.}~\bibnamefont {Cothrine}}, \bibinfo {author}
  {\bibfnamefont {K.}~\bibnamefont {Watanabe}}, \bibinfo {author}
  {\bibfnamefont {T.}~\bibnamefont {Taniguchi}}, \bibinfo {author}
  {\bibfnamefont {D.}~\bibnamefont {Mandrus}}, \bibinfo {author} {\bibfnamefont
  {K.}~\bibnamefont {Barmak}}, \bibinfo {author} {\bibfnamefont
  {J.}~\bibnamefont {Hone}}, \bibinfo {author} {\bibfnamefont {A.~J.}\
  \bibnamefont {Millis}}, \bibinfo {author} {\bibfnamefont {A.~N.}\
  \bibnamefont {Pasupathy}},\ and\ \bibinfo {author} {\bibfnamefont {C.~R.}\
  \bibnamefont {Dean}},\ }\href {https://arxiv.org/abs/2406.03418} {\bibinfo
  {title} {Superconductivity in twisted bilayer wse$_2$}} (\bibinfo {year}
  {2024}),\ \Eprint {https://arxiv.org/abs/2406.03418} {arXiv:2406.03418
  [cond-mat.mes-hall]} \BibitemShut {NoStop}%
\bibitem [{\citenamefont {Wu}\ \emph {et~al.}(2018)\citenamefont {Wu},
  \citenamefont {Lovorn}, \citenamefont {Tutuc},\ and\ \citenamefont
  {MacDonald}}]{MacDonald_TMD2018}%
  \BibitemOpen
  \bibfield  {author} {\bibinfo {author} {\bibfnamefont {F.}~\bibnamefont
  {Wu}}, \bibinfo {author} {\bibfnamefont {T.}~\bibnamefont {Lovorn}}, \bibinfo
  {author} {\bibfnamefont {E.}~\bibnamefont {Tutuc}},\ and\ \bibinfo {author}
  {\bibfnamefont {A.~H.}\ \bibnamefont {MacDonald}},\ }\bibfield  {title}
  {\bibinfo {title} {Hubbard model physics in transition metal dichalcogenide
  moir\'e bands},\ }\href {https://doi.org/10.1103/PhysRevLett.121.026402}
  {\bibfield  {journal} {\bibinfo  {journal} {Phys. Rev. Lett.}\ }\textbf
  {\bibinfo {volume} {121}},\ \bibinfo {pages} {026402} (\bibinfo {year}
  {2018})}\BibitemShut {NoStop}%
\bibitem [{\citenamefont {Pan}\ \emph {et~al.}(2020)\citenamefont {Pan},
  \citenamefont {Wu},\ and\ \citenamefont {Das~Sarma}}]{DasSarma_TMD2020}%
  \BibitemOpen
  \bibfield  {author} {\bibinfo {author} {\bibfnamefont {H.}~\bibnamefont
  {Pan}}, \bibinfo {author} {\bibfnamefont {F.}~\bibnamefont {Wu}},\ and\
  \bibinfo {author} {\bibfnamefont {S.}~\bibnamefont {Das~Sarma}},\ }\bibfield
  {title} {\bibinfo {title} {Band topology, hubbard model, heisenberg model,
  and dzyaloshinskii-moriya interaction in twisted bilayer
  ${\mathrm{wse}}_{2}$},\ }\href
  {https://doi.org/10.1103/PhysRevResearch.2.033087} {\bibfield  {journal}
  {\bibinfo  {journal} {Phys. Rev. Res.}\ }\textbf {\bibinfo {volume} {2}},\
  \bibinfo {pages} {033087} (\bibinfo {year} {2020})}\BibitemShut {NoStop}%
\bibitem [{\citenamefont {Zang}\ \emph {et~al.}(2021)\citenamefont {Zang},
  \citenamefont {Wang}, \citenamefont {Cano},\ and\ \citenamefont
  {Millis}}]{Millis_TMD2021}%
  \BibitemOpen
  \bibfield  {author} {\bibinfo {author} {\bibfnamefont {J.}~\bibnamefont
  {Zang}}, \bibinfo {author} {\bibfnamefont {J.}~\bibnamefont {Wang}}, \bibinfo
  {author} {\bibfnamefont {J.}~\bibnamefont {Cano}},\ and\ \bibinfo {author}
  {\bibfnamefont {A.~J.}\ \bibnamefont {Millis}},\ }\bibfield  {title}
  {\bibinfo {title} {Hartree-fock study of the moir\'e hubbard model for
  twisted bilayer transition metal dichalcogenides},\ }\href
  {https://doi.org/10.1103/PhysRevB.104.075150} {\bibfield  {journal} {\bibinfo
   {journal} {Phys. Rev. B}\ }\textbf {\bibinfo {volume} {104}},\ \bibinfo
  {pages} {075150} (\bibinfo {year} {2021})}\BibitemShut {NoStop}%
\bibitem [{\citenamefont {Devakul}\ \emph {et~al.}(2021)\citenamefont
  {Devakul}, \citenamefont {Cr{\'e}pel}, \citenamefont {Zhang},\ and\
  \citenamefont {Fu}}]{Devakul:2021kn}%
  \BibitemOpen
  \bibfield  {author} {\bibinfo {author} {\bibfnamefont {T.}~\bibnamefont
  {Devakul}}, \bibinfo {author} {\bibfnamefont {V.}~\bibnamefont {Cr{\'e}pel}},
  \bibinfo {author} {\bibfnamefont {Y.}~\bibnamefont {Zhang}},\ and\ \bibinfo
  {author} {\bibfnamefont {L.}~\bibnamefont {Fu}},\ }\bibfield  {title}
  {\bibinfo {title} {Magic in twisted transition metal dichalcogenide
  bilayers},\ }\href {https://doi.org/10.1038/s41467-021-27042-9} {\bibfield
  {journal} {\bibinfo  {journal} {Nature Communications}\ }\textbf {\bibinfo
  {volume} {12}},\ \bibinfo {pages} {6730} (\bibinfo {year}
  {2021})}\BibitemShut {NoStop}%
\bibitem [{\citenamefont {Christos}\ \emph {et~al.}(2024)\citenamefont
  {Christos}, \citenamefont {Bonetti},\ and\ \citenamefont
  {Scheurer}}]{Scheurer2024}%
  \BibitemOpen
  \bibfield  {author} {\bibinfo {author} {\bibfnamefont {M.}~\bibnamefont
  {Christos}}, \bibinfo {author} {\bibfnamefont {P.~M.}\ \bibnamefont
  {Bonetti}},\ and\ \bibinfo {author} {\bibfnamefont {M.~S.}\ \bibnamefont
  {Scheurer}},\ }\href {https://arxiv.org/abs/2407.02393} {\bibinfo {title}
  {Approximate symmetries, insulators, and superconductivity in continuum-model
  description of twisted wse$_2$}} (\bibinfo {year} {2024}),\ \Eprint
  {https://arxiv.org/abs/2407.02393} {arXiv:2407.02393 [cond-mat.supr-con]}
  \BibitemShut {NoStop}%
\bibitem [{\citenamefont {Wilhelm}\ \emph {et~al.}(2024)\citenamefont
  {Wilhelm}, \citenamefont {Läuchli},\ and\ \citenamefont
  {Scheurer}}]{Scheurer2024_1}%
  \BibitemOpen
  \bibfield  {author} {\bibinfo {author} {\bibfnamefont {P.~H.}\ \bibnamefont
  {Wilhelm}}, \bibinfo {author} {\bibfnamefont {A.~M.}\ \bibnamefont
  {Läuchli}},\ and\ \bibinfo {author} {\bibfnamefont {M.~S.}\ \bibnamefont
  {Scheurer}},\ }\href {https://arxiv.org/abs/2406.09505} {\bibinfo {title} {A
  novel perspective on ideal chern bands with strong short-range repulsion:
  Applications to correlated metals, superconductivity, and topological order}}
  (\bibinfo {year} {2024}),\ \Eprint {https://arxiv.org/abs/2406.09505}
  {arXiv:2406.09505 [cond-mat.str-el]} \BibitemShut {NoStop}%
\bibitem [{\citenamefont {Guerci}\ \emph {et~al.}(2024)\citenamefont {Guerci},
  \citenamefont {Kaplan}, \citenamefont {Ingham}, \citenamefont {Pixley},\ and\
  \citenamefont {Millis}}]{Guerci2024}%
  \BibitemOpen
  \bibfield  {author} {\bibinfo {author} {\bibfnamefont {D.}~\bibnamefont
  {Guerci}}, \bibinfo {author} {\bibfnamefont {D.}~\bibnamefont {Kaplan}},
  \bibinfo {author} {\bibfnamefont {J.}~\bibnamefont {Ingham}}, \bibinfo
  {author} {\bibfnamefont {J.~H.}\ \bibnamefont {Pixley}},\ and\ \bibinfo
  {author} {\bibfnamefont {A.~J.}\ \bibnamefont {Millis}},\ }\href
  {https://arxiv.org/abs/2408.16075} {\bibinfo {title} {Topological
  superconductivity from repulsive interactions in twisted wse$_2$}} (\bibinfo
  {year} {2024}),\ \Eprint {https://arxiv.org/abs/2408.16075} {arXiv:2408.16075
  [cond-mat.supr-con]} \BibitemShut {NoStop}%
\bibitem [{\citenamefont {Aji}\ and\ \citenamefont {Varma}(2007)}]{Aji-V-qcf1}%
  \BibitemOpen
  \bibfield  {author} {\bibinfo {author} {\bibfnamefont {V.}~\bibnamefont
  {Aji}}\ and\ \bibinfo {author} {\bibfnamefont {C.~M.}\ \bibnamefont
  {Varma}},\ }\bibfield  {title} {\bibinfo {title} {Theory of the quantum
  critical fluctuations in cuprate superconductors},\ }\href
  {https://doi.org/10.1103/PhysRevLett.99.067003} {\bibfield  {journal}
  {\bibinfo  {journal} {Phys. Rev. Lett.}\ }\textbf {\bibinfo {volume} {99}},\
  \bibinfo {pages} {067003} (\bibinfo {year} {2007})}\BibitemShut {NoStop}%
\bibitem [{\citenamefont {Varma}(2016)}]{Varma_IOPrev2016}%
  \BibitemOpen
  \bibfield  {author} {\bibinfo {author} {\bibfnamefont {C.~M.}\ \bibnamefont
  {Varma}},\ }\bibfield  {title} {\bibinfo {title} {Quantum-critical
  fluctuations in 2d metals: strange metals and superconductivity in
  antiferromagnets and in cuprates},\ }\href
  {https://doi.org/10.1088/0034-4885/79/8/082501} {\bibfield  {journal}
  {\bibinfo  {journal} {Reports on Progress in Physics}\ }\textbf {\bibinfo
  {volume} {79}},\ \bibinfo {pages} {082501} (\bibinfo {year}
  {2016})}\BibitemShut {NoStop}%
\bibitem [{\citenamefont {Varma}\ \emph {et~al.}(1989)\citenamefont {Varma},
  \citenamefont {Littlewood}, \citenamefont {Schmitt-Rink}, \citenamefont
  {Abrahams},\ and\ \citenamefont {Ruckenstein}}]{CMV-MFL}%
  \BibitemOpen
  \bibfield  {author} {\bibinfo {author} {\bibfnamefont {C.~M.}\ \bibnamefont
  {Varma}}, \bibinfo {author} {\bibfnamefont {P.~B.}\ \bibnamefont
  {Littlewood}}, \bibinfo {author} {\bibfnamefont {S.}~\bibnamefont
  {Schmitt-Rink}}, \bibinfo {author} {\bibfnamefont {E.}~\bibnamefont
  {Abrahams}},\ and\ \bibinfo {author} {\bibfnamefont {A.~E.}\ \bibnamefont
  {Ruckenstein}},\ }\bibfield  {title} {\bibinfo {title} {Phenomenology of the
  normal state of copper oxide high-temperature superconductors},\ }\href
  {https://doi.org/10.1103/PhysRevLett.63.1996} {\bibfield  {journal} {\bibinfo
   {journal} {Phys. Rev. Lett.}\ }\textbf {\bibinfo {volume} {63}},\ \bibinfo
  {pages} {1996} (\bibinfo {year} {1989})}\BibitemShut {NoStop}%
\bibitem [{\citenamefont {Kaminski}\ \emph {et~al.}(2002)\citenamefont
  {Kaminski}, \citenamefont {Rosenkranz}, \citenamefont {Fretwell},
  \citenamefont {Campuzano}, \citenamefont {Li}, \citenamefont {Raffy},
  \citenamefont {Cullen}, \citenamefont {You}, \citenamefont {Olson},
  \citenamefont {Varma},\ and\ \citenamefont {H\"{o}chst}}]{Kaminski-diARPES}%
  \BibitemOpen
  \bibfield  {author} {\bibinfo {author} {\bibfnamefont {A.}~\bibnamefont
  {Kaminski}}, \bibinfo {author} {\bibfnamefont {S.}~\bibnamefont
  {Rosenkranz}}, \bibinfo {author} {\bibfnamefont {H.~M.}\ \bibnamefont
  {Fretwell}}, \bibinfo {author} {\bibfnamefont {J.~C.}\ \bibnamefont
  {Campuzano}}, \bibinfo {author} {\bibfnamefont {Z.}~\bibnamefont {Li}},
  \bibinfo {author} {\bibfnamefont {H.}~\bibnamefont {Raffy}}, \bibinfo
  {author} {\bibfnamefont {W.~G.}\ \bibnamefont {Cullen}}, \bibinfo {author}
  {\bibfnamefont {H.}~\bibnamefont {You}}, \bibinfo {author} {\bibfnamefont
  {C.~G.}\ \bibnamefont {Olson}}, \bibinfo {author} {\bibfnamefont {C.~M.}\
  \bibnamefont {Varma}},\ and\ \bibinfo {author} {\bibfnamefont
  {H.}~\bibnamefont {H\"{o}chst}},\ }\bibfield  {title} {\bibinfo {title}
  {{Spontaneous breaking of time-reversal symmetry in the pseudogap state of a
  high-Tc superconductor.}},\ }\href {https://doi.org/10.1038/416610a}
  {\bibfield  {journal} {\bibinfo  {journal} {Nature}\ }\textbf {\bibinfo
  {volume} {416}},\ \bibinfo {pages} {610} (\bibinfo {year}
  {2002})}\BibitemShut {NoStop}%
\bibitem [{\citenamefont {Simon}\ and\ \citenamefont
  {Varma}(2002)}]{simon-cmv}%
  \BibitemOpen
  \bibfield  {author} {\bibinfo {author} {\bibfnamefont {M.~E.}\ \bibnamefont
  {Simon}}\ and\ \bibinfo {author} {\bibfnamefont {C.~M.}\ \bibnamefont
  {Varma}},\ }\bibfield  {title} {\bibinfo {title} {Detection and implications
  of a time-reversal breaking state in underdoped cuprates},\ }\href
  {https://doi.org/10.1103/PhysRevLett.89.247003} {\bibfield  {journal}
  {\bibinfo  {journal} {Phys. Rev. Lett.}\ }\textbf {\bibinfo {volume} {89}},\
  \bibinfo {pages} {247003} (\bibinfo {year} {2002})}\BibitemShut {NoStop}%
\bibitem [{\citenamefont {Bourges}\ and\ \citenamefont
  {Sidis}(2011)}]{Bourges-rev}%
  \BibitemOpen
  \bibfield  {author} {\bibinfo {author} {\bibfnamefont {P.}~\bibnamefont
  {Bourges}}\ and\ \bibinfo {author} {\bibfnamefont {Y.}~\bibnamefont
  {Sidis}},\ }\bibfield  {title} {\bibinfo {title} {Novel magnetic order in the
  pseudogap state of high- copper oxides superconductors},\ }\href
  {https://doi.org/http://dx.doi.org/10.1016/j.crhy.2011.04.006} {\bibfield
  {journal} {\bibinfo  {journal} {Comptes Rendus Physique}\ }\textbf {\bibinfo
  {volume} {12}},\ \bibinfo {pages} {461 } (\bibinfo {year}
  {2011})}\BibitemShut {NoStop}%
\bibitem [{\citenamefont {von Lohneysen}(1996)}]{Lohneysen1996}%
  \BibitemOpen
  \bibfield  {author} {\bibinfo {author} {\bibfnamefont {H.}~\bibnamefont {von
  Lohneysen}},\ }\bibfield  {title} {\bibinfo {title} {Non-fermi-liquid
  behaviour in the heavy-fermion system},\ }\href
  {https://doi.org/10.1088/0953-8984/8/48/003} {\bibfield  {journal} {\bibinfo
  {journal} {Journal of Physics: Condensed Matter}\ }\textbf {\bibinfo {volume}
  {8}},\ \bibinfo {pages} {9689} (\bibinfo {year} {1996})}\BibitemShut
  {NoStop}%
\bibitem [{\citenamefont {Shibauchi}\ \emph {et~al.}(2014)\citenamefont
  {Shibauchi}, \citenamefont {Carrington},\ and\ \citenamefont
  {Matsuda}}]{ShibauchiQCP}%
  \BibitemOpen
  \bibfield  {author} {\bibinfo {author} {\bibfnamefont {T.}~\bibnamefont
  {Shibauchi}}, \bibinfo {author} {\bibfnamefont {A.}~\bibnamefont
  {Carrington}},\ and\ \bibinfo {author} {\bibfnamefont {Y.}~\bibnamefont
  {Matsuda}},\ }\bibfield  {title} {\bibinfo {title} {A quantum critical point
  lying beneath the superconducting dome in iron pnictides},\ }\href
  {https://doi.org/10.1146/annurev-conmatphys-031113-133921} {\bibfield
  {journal} {\bibinfo  {journal} {Annual Review of Condensed Matter Physics}\
  }\textbf {\bibinfo {volume} {5}},\ \bibinfo {pages} {113} (\bibinfo {year}
  {2014})}\BibitemShut {NoStop}%
\bibitem [{\citenamefont {Sachdev}(2011)}]{Sachdev_2011}%
  \BibitemOpen
  \bibfield  {author} {\bibinfo {author} {\bibfnamefont {S.}~\bibnamefont
  {Sachdev}},\ }\href@noop {} {\emph {\bibinfo {title} {Quantum Phase
  Transitions}}},\ \bibinfo {edition} {2nd}\ ed.\ (\bibinfo  {publisher}
  {Cambridge University Press},\ \bibinfo {year} {2011})\BibitemShut {NoStop}%
\bibitem [{\citenamefont {Varma}\ \emph {et~al.}(2002)\citenamefont {Varma},
  \citenamefont {Nussinov},\ and\ \citenamefont {van Saarloos}}]{VNS}%
  \BibitemOpen
  \bibfield  {author} {\bibinfo {author} {\bibfnamefont {C.~M.}\ \bibnamefont
  {Varma}}, \bibinfo {author} {\bibfnamefont {Z.}~\bibnamefont {Nussinov}},\
  and\ \bibinfo {author} {\bibfnamefont {W.}~\bibnamefont {van Saarloos}},\
  }\bibfield  {title} {\bibinfo {title} {Singular or non-fermi liquids},\
  }\href@noop {} {\bibfield  {journal} {\bibinfo  {journal} {Phys. Reports}\ }
  (\bibinfo {year} {2002})}\BibitemShut {NoStop}%
\bibitem [{\citenamefont {Taillefer}(2010)}]{TailleferREV2010}%
  \BibitemOpen
  \bibfield  {author} {\bibinfo {author} {\bibfnamefont {L.}~\bibnamefont
  {Taillefer}},\ }\bibfield  {title} {\bibinfo {title} {Scattering and pairing
  in cuprate superconductors},\ }\href
  {https://doi.org/https://doi.org/10.1146/annurev-conmatphys-070909-104117}
  {\bibfield  {journal} {\bibinfo  {journal} {Annual Review of Condensed Matter
  Physics}\ }\textbf {\bibinfo {volume} {1}},\ \bibinfo {pages} {51} (\bibinfo
  {year} {2010})}\BibitemShut {NoStop}%
\bibitem [{\citenamefont {Klein}\ \emph {et~al.}(2020)\citenamefont {Klein},
  \citenamefont {Chubukov}, \citenamefont {Schattner},\ and\ \citenamefont
  {Berg}}]{klein2020}%
  \BibitemOpen
  \bibfield  {author} {\bibinfo {author} {\bibfnamefont {A.}~\bibnamefont
  {Klein}}, \bibinfo {author} {\bibfnamefont {A.~V.}\ \bibnamefont {Chubukov}},
  \bibinfo {author} {\bibfnamefont {Y.}~\bibnamefont {Schattner}},\ and\
  \bibinfo {author} {\bibfnamefont {E.}~\bibnamefont {Berg}},\ }\bibfield
  {title} {\bibinfo {title} {Normal state properties of quantum critical metals
  at finite temperature},\ }\href {https://doi.org/10.1103/PhysRevX.10.031053}
  {\bibfield  {journal} {\bibinfo  {journal} {Phys. Rev. X}\ }\textbf {\bibinfo
  {volume} {10}},\ \bibinfo {pages} {031053} (\bibinfo {year}
  {2020})}\BibitemShut {NoStop}%
\bibitem [{\citenamefont {Lee}(2021)}]{palee2021}%
  \BibitemOpen
  \bibfield  {author} {\bibinfo {author} {\bibfnamefont {P.~A.}\ \bibnamefont
  {Lee}},\ }\bibfield  {title} {\bibinfo {title} {Low-temperature $t$-linear
  resistivity due to umklapp scattering from a critical mode},\ }\href
  {https://doi.org/10.1103/PhysRevB.104.035140} {\bibfield  {journal} {\bibinfo
   {journal} {Phys. Rev. B}\ }\textbf {\bibinfo {volume} {104}},\ \bibinfo
  {pages} {035140} (\bibinfo {year} {2021})}\BibitemShut {NoStop}%
\bibitem [{\citenamefont {Zang}\ \emph {et~al.}(2022)\citenamefont {Zang},
  \citenamefont {Wang}, \citenamefont {Cano}, \citenamefont {Georges},\ and\
  \citenamefont {Millis}}]{Cano2022}%
  \BibitemOpen
  \bibfield  {author} {\bibinfo {author} {\bibfnamefont {J.}~\bibnamefont
  {Zang}}, \bibinfo {author} {\bibfnamefont {J.}~\bibnamefont {Wang}}, \bibinfo
  {author} {\bibfnamefont {J.}~\bibnamefont {Cano}}, \bibinfo {author}
  {\bibfnamefont {A.}~\bibnamefont {Georges}},\ and\ \bibinfo {author}
  {\bibfnamefont {A.~J.}\ \bibnamefont {Millis}},\ }\bibfield  {title}
  {\bibinfo {title} {Dynamical mean-field theory of moir\'e bilayer transition
  metal dichalcogenides: Phase diagram, resistivity, and quantum criticality},\
  }\href {https://doi.org/10.1103/PhysRevX.12.021064} {\bibfield  {journal}
  {\bibinfo  {journal} {Phys. Rev. X}\ }\textbf {\bibinfo {volume} {12}},\
  \bibinfo {pages} {021064} (\bibinfo {year} {2022})}\BibitemShut {NoStop}%
\bibitem [{\citenamefont {Patel}\ \emph {et~al.}(2023)\citenamefont {Patel},
  \citenamefont {Guo}, \citenamefont {Esterlis},\ and\ \citenamefont
  {Sachdev}}]{Sachdev_science2023}%
  \BibitemOpen
  \bibfield  {author} {\bibinfo {author} {\bibfnamefont {A.~A.}\ \bibnamefont
  {Patel}}, \bibinfo {author} {\bibfnamefont {H.}~\bibnamefont {Guo}}, \bibinfo
  {author} {\bibfnamefont {I.}~\bibnamefont {Esterlis}},\ and\ \bibinfo
  {author} {\bibfnamefont {S.}~\bibnamefont {Sachdev}},\ }\bibfield  {title}
  {\bibinfo {title} {Universal theory of strange metals from spatially random
  interactions},\ }\href@noop {} {\bibfield  {journal} {\bibinfo  {journal}
  {Science}\ }\textbf {\bibinfo {volume} {381}},\ \bibinfo {pages} {790}
  (\bibinfo {year} {2023})}\BibitemShut {NoStop}%
\bibitem [{\citenamefont {Shtyk}\ \emph {et~al.}(2017)\citenamefont {Shtyk},
  \citenamefont {Goldstein},\ and\ \citenamefont {Chamon}}]{Shtyk2017}%
  \BibitemOpen
  \bibfield  {author} {\bibinfo {author} {\bibfnamefont {A.}~\bibnamefont
  {Shtyk}}, \bibinfo {author} {\bibfnamefont {G.}~\bibnamefont {Goldstein}},\
  and\ \bibinfo {author} {\bibfnamefont {C.}~\bibnamefont {Chamon}},\
  }\bibfield  {title} {\bibinfo {title} {{Electrons at the monkey saddle: A
  multicritical Lifshitz point}},\ }\href
  {https://doi.org/10.1103/PhysRevB.95.035137} {\bibfield  {journal} {\bibinfo
  {journal} {Phys. Rev. B}\ }\textbf {\bibinfo {volume} {95}},\ \bibinfo
  {pages} {035137} (\bibinfo {year} {2017})}\BibitemShut {NoStop}%
\bibitem [{\citenamefont {Classen}\ \emph {et~al.}(2020)\citenamefont
  {Classen}, \citenamefont {Chubukov}, \citenamefont {Honerkamp},\ and\
  \citenamefont {Scherer}}]{Classen2020PRB}%
  \BibitemOpen
  \bibfield  {author} {\bibinfo {author} {\bibfnamefont {L.}~\bibnamefont
  {Classen}}, \bibinfo {author} {\bibfnamefont {A.~V.}\ \bibnamefont
  {Chubukov}}, \bibinfo {author} {\bibfnamefont {C.}~\bibnamefont
  {Honerkamp}},\ and\ \bibinfo {author} {\bibfnamefont {M.~M.}\ \bibnamefont
  {Scherer}},\ }\bibfield  {title} {\bibinfo {title} {Competing orders at
  higher-order van hove points},\ }\href
  {https://doi.org/10.1103/PhysRevB.102.125141} {\bibfield  {journal} {\bibinfo
   {journal} {Phys. Rev. B}\ }\textbf {\bibinfo {volume} {102}},\ \bibinfo
  {pages} {125141} (\bibinfo {year} {2020})}\BibitemShut {NoStop}%
\bibitem [{\citenamefont {Classen}\ and\ \citenamefont
  {Betouras}(2024)}]{Classen2024}%
  \BibitemOpen
  \bibfield  {author} {\bibinfo {author} {\bibfnamefont {L.}~\bibnamefont
  {Classen}}\ and\ \bibinfo {author} {\bibfnamefont {J.~J.}\ \bibnamefont
  {Betouras}},\ }\href {https://arxiv.org/abs/2405.20226} {\bibinfo {title}
  {High-order van hove singularities and their connection to flat bands}}
  (\bibinfo {year} {2024}),\ \Eprint {https://arxiv.org/abs/2405.20226}
  {arXiv:2405.20226 [cond-mat.str-el]} \BibitemShut {NoStop}%
\bibitem [{\citenamefont {Isobe}\ and\ \citenamefont {Fu}(2019)}]{Isobe2019}%
  \BibitemOpen
  \bibfield  {author} {\bibinfo {author} {\bibfnamefont {H.}~\bibnamefont
  {Isobe}}\ and\ \bibinfo {author} {\bibfnamefont {L.}~\bibnamefont {Fu}},\
  }\bibfield  {title} {\bibinfo {title} {Supermetal},\ }\href
  {https://doi.org/10.1103/PhysRevResearch.1.033206} {\bibfield  {journal}
  {\bibinfo  {journal} {Phys. Rev. Res.}\ }\textbf {\bibinfo {volume} {1}},\
  \bibinfo {pages} {033206} (\bibinfo {year} {2019})}\BibitemShut {NoStop}%
\bibitem [{\citenamefont {Chichinadze}\ \emph {et~al.}(2022)\citenamefont
  {Chichinadze}, \citenamefont {Classen}, \citenamefont {Wang},\ and\
  \citenamefont {Chubukov}}]{Classen}%
  \BibitemOpen
  \bibfield  {author} {\bibinfo {author} {\bibfnamefont {D.~V.}\ \bibnamefont
  {Chichinadze}}, \bibinfo {author} {\bibfnamefont {L.}~\bibnamefont
  {Classen}}, \bibinfo {author} {\bibfnamefont {Y.}~\bibnamefont {Wang}},\ and\
  \bibinfo {author} {\bibfnamefont {A.~V.}\ \bibnamefont {Chubukov}},\
  }\bibfield  {title} {\bibinfo {title} {Cascade of transitions in twisted and
  non-twisted graphene layers within the van hove scenario},\ }\href
  {https://doi.org/10.1038/s41535-022-00520-z} {\bibfield  {journal} {\bibinfo
  {journal} {npj Quantum Materials}\ }\textbf {\bibinfo {volume} {7}},\
  \bibinfo {pages} {114} (\bibinfo {year} {2022})}\BibitemShut {NoStop}%
\bibitem [{\citenamefont {Ojajärvi}\ \emph {et~al.}(2024)\citenamefont
  {Ojajärvi}, \citenamefont {Chubukov}, \citenamefont {Lee}, \citenamefont
  {Garst},\ and\ \citenamefont {Schmalian}}]{Schmalian}%
  \BibitemOpen
  \bibfield  {author} {\bibinfo {author} {\bibfnamefont {R.}~\bibnamefont
  {Ojajärvi}}, \bibinfo {author} {\bibfnamefont {A.~V.}\ \bibnamefont
  {Chubukov}}, \bibinfo {author} {\bibfnamefont {Y.-C.}\ \bibnamefont {Lee}},
  \bibinfo {author} {\bibfnamefont {M.}~\bibnamefont {Garst}},\ and\ \bibinfo
  {author} {\bibfnamefont {J.}~\bibnamefont {Schmalian}},\ }\href
  {https://arxiv.org/abs/2408.05572} {\bibinfo {title} {Pairing at a single van
  hove point}} (\bibinfo {year} {2024}),\ \Eprint
  {https://arxiv.org/abs/2408.05572} {arXiv:2408.05572 [cond-mat.supr-con]}
  \BibitemShut {NoStop}%
\bibitem [{Note1()}]{Note1}%
  \BibitemOpen
  \bibinfo {note} {See ~\cite
  {KL,Raghu2010,kagan,*Chubukov93,Guinea2024,Schmalian} and references therein
  on KL scenario.}\BibitemShut {Stop}%
\bibitem [{\citenamefont {Nandkishore}\ \emph {et~al.}(2012)\citenamefont
  {Nandkishore}, \citenamefont {Levitov},\ and\ \citenamefont
  {Chubukov}}]{Nandkishore2012}%
  \BibitemOpen
  \bibfield  {author} {\bibinfo {author} {\bibfnamefont {R.}~\bibnamefont
  {Nandkishore}}, \bibinfo {author} {\bibfnamefont {L.~S.}\ \bibnamefont
  {Levitov}},\ and\ \bibinfo {author} {\bibfnamefont {A.~V.}\ \bibnamefont
  {Chubukov}},\ }\bibfield  {title} {\bibinfo {title} {Chiral superconductivity
  from repulsive interactions in doped graphene},\ }\href@noop {} {\bibfield
  {journal} {\bibinfo  {journal} {Nature Physics}\ }\textbf {\bibinfo {volume}
  {8}},\ \bibinfo {pages} {158} (\bibinfo {year} {2012})}\BibitemShut {NoStop}%
\bibitem [{\citenamefont {Dong}\ \emph {et~al.}(2023)\citenamefont {Dong},
  \citenamefont {Levitov},\ and\ \citenamefont {Chubukov}}]{Dong2024}%
  \BibitemOpen
  \bibfield  {author} {\bibinfo {author} {\bibfnamefont {Z.}~\bibnamefont
  {Dong}}, \bibinfo {author} {\bibfnamefont {L.}~\bibnamefont {Levitov}},\ and\
  \bibinfo {author} {\bibfnamefont {A.~V.}\ \bibnamefont {Chubukov}},\
  }\bibfield  {title} {\bibinfo {title} {Superconductivity near spin and valley
  orders in graphene multilayers},\ }\href
  {https://doi.org/10.1103/PhysRevB.108.134503} {\bibfield  {journal} {\bibinfo
   {journal} {Phys. Rev. B}\ }\textbf {\bibinfo {volume} {108}},\ \bibinfo
  {pages} {134503} (\bibinfo {year} {2023})}\BibitemShut {NoStop}%
\bibitem [{\citenamefont {Miyake}\ \emph {et~al.}(1986)\citenamefont {Miyake},
  \citenamefont {Schmitt-Rink},\ and\ \citenamefont {Varma}}]{MiyakeSV1986}%
  \BibitemOpen
  \bibfield  {author} {\bibinfo {author} {\bibfnamefont {K.}~\bibnamefont
  {Miyake}}, \bibinfo {author} {\bibfnamefont {S.}~\bibnamefont
  {Schmitt-Rink}},\ and\ \bibinfo {author} {\bibfnamefont {C.~M.}\ \bibnamefont
  {Varma}},\ }\bibfield  {title} {\bibinfo {title} {Spin-fluctuation-mediated
  even-parity pairing in heavy-fermion superconductors},\ }\href
  {https://doi.org/10.1103/PhysRevB.34.6554} {\bibfield  {journal} {\bibinfo
  {journal} {Phys. Rev. B}\ }\textbf {\bibinfo {volume} {34}},\ \bibinfo
  {pages} {6554} (\bibinfo {year} {1986})}\BibitemShut {NoStop}%
\bibitem [{\citenamefont {Scalapino}\ \emph {et~al.}(1986)\citenamefont
  {Scalapino}, \citenamefont {Loh},\ and\ \citenamefont
  {Hirsch}}]{Scalapino1986}%
  \BibitemOpen
  \bibfield  {author} {\bibinfo {author} {\bibfnamefont {D.~J.}\ \bibnamefont
  {Scalapino}}, \bibinfo {author} {\bibfnamefont {E.}~\bibnamefont {Loh}},\
  and\ \bibinfo {author} {\bibfnamefont {J.~E.}\ \bibnamefont {Hirsch}},\
  }\bibfield  {title} {\bibinfo {title} {$d$-wave pairing near a
  spin-density-wave instability},\ }\href
  {https://doi.org/10.1103/PhysRevB.34.8190} {\bibfield  {journal} {\bibinfo
  {journal} {Phys. Rev. B}\ }\textbf {\bibinfo {volume} {34}},\ \bibinfo
  {pages} {8190} (\bibinfo {year} {1986})}\BibitemShut {NoStop}%
\bibitem [{\citenamefont {Varma}(2020)}]{VarmaRMP2020}%
  \BibitemOpen
  \bibfield  {author} {\bibinfo {author} {\bibfnamefont {C.~M.}\ \bibnamefont
  {Varma}},\ }\bibfield  {title} {\bibinfo {title} {Colloquium: Linear in
  temperature resistivity and associated mysteries including high temperature
  superconductivity},\ }\href {https://doi.org/10.1103/RevModPhys.92.031001}
  {\bibfield  {journal} {\bibinfo  {journal} {Rev. Mod. Phys.}\ }\textbf
  {\bibinfo {volume} {92}},\ \bibinfo {pages} {031001} (\bibinfo {year}
  {2020})}\BibitemShut {NoStop}%
\bibitem [{\citenamefont {Zhu}\ \emph {et~al.}(2015)\citenamefont {Zhu},
  \citenamefont {Chen},\ and\ \citenamefont {Varma}}]{ZhuChenV2015}%
  \BibitemOpen
  \bibfield  {author} {\bibinfo {author} {\bibfnamefont {L.}~\bibnamefont
  {Zhu}}, \bibinfo {author} {\bibfnamefont {Y.}~\bibnamefont {Chen}},\ and\
  \bibinfo {author} {\bibfnamefont {C.~M.}\ \bibnamefont {Varma}},\ }\bibfield
  {title} {\bibinfo {title} {Local quantum criticality in the two-dimensional
  dissipative quantum xy model},\ }\href
  {https://doi.org/10.1103/PhysRevB.91.205129} {\bibfield  {journal} {\bibinfo
  {journal} {Phys. Rev. B}\ }\textbf {\bibinfo {volume} {91}},\ \bibinfo
  {pages} {205129} (\bibinfo {year} {2015})}\BibitemShut {NoStop}%
\bibitem [{\citenamefont {Zhu}\ \emph {et~al.}(2016)\citenamefont {Zhu},
  \citenamefont {Hou},\ and\ \citenamefont {Varma}}]{ZhuHouV2016}%
  \BibitemOpen
  \bibfield  {author} {\bibinfo {author} {\bibfnamefont {L.}~\bibnamefont
  {Zhu}}, \bibinfo {author} {\bibfnamefont {C.}~\bibnamefont {Hou}},\ and\
  \bibinfo {author} {\bibfnamefont {C.~M.}\ \bibnamefont {Varma}},\ }\bibfield
  {title} {\bibinfo {title} {Quantum criticality in the two-dimensional
  dissipative quantum xy model},\ }\href
  {https://doi.org/10.1103/PhysRevB.94.235156} {\bibfield  {journal} {\bibinfo
  {journal} {Phys. Rev. B}\ }\textbf {\bibinfo {volume} {94}},\ \bibinfo
  {pages} {235156} (\bibinfo {year} {2016})}\BibitemShut {NoStop}%
\bibitem [{\citenamefont {Hou}\ and\ \citenamefont {Varma}(2016)}]{Hou-CMV-RG}%
  \BibitemOpen
  \bibfield  {author} {\bibinfo {author} {\bibfnamefont {C.}~\bibnamefont
  {Hou}}\ and\ \bibinfo {author} {\bibfnamefont {C.~M.}\ \bibnamefont
  {Varma}},\ }\bibfield  {title} {\bibinfo {title} {Phase diagram and
  quantum-criticality of the two dimensional dissipative quantum xy model},\
  }\href@noop {} {\bibfield  {journal} {\bibinfo  {journal} {Unpublished}\ }
  (\bibinfo {year} {2016})}\BibitemShut {NoStop}%
\bibitem [{\citenamefont {Schroeder}\ \emph {et~al.}(2000)\citenamefont
  {Schroeder}, \citenamefont {Aeppli}, \citenamefont {Coldea}, \citenamefont
  {Adams}, \citenamefont {Stockert}, \citenamefont {Lohneysen}, \citenamefont
  {Bucher}, \citenamefont {Ramazashvili},\ and\ \citenamefont
  {Coleman}}]{Schroeder2000}%
  \BibitemOpen
  \bibfield  {author} {\bibinfo {author} {\bibfnamefont {A.}~\bibnamefont
  {Schroeder}}, \bibinfo {author} {\bibfnamefont {G.}~\bibnamefont {Aeppli}},
  \bibinfo {author} {\bibfnamefont {R.}~\bibnamefont {Coldea}}, \bibinfo
  {author} {\bibfnamefont {M.}~\bibnamefont {Adams}}, \bibinfo {author}
  {\bibfnamefont {O.}~\bibnamefont {Stockert}}, \bibinfo {author}
  {\bibfnamefont {H.~v.}\ \bibnamefont {Lohneysen}}, \bibinfo {author}
  {\bibfnamefont {E.}~\bibnamefont {Bucher}}, \bibinfo {author} {\bibfnamefont
  {R.}~\bibnamefont {Ramazashvili}},\ and\ \bibinfo {author} {\bibfnamefont
  {P.}~\bibnamefont {Coleman}},\ }\bibfield  {title} {\bibinfo {title} {Onset
  of antiferromagnetism in heavy-fermion metals},\ }\href
  {https://doi.org/10.1038/35030039} {\bibfield  {journal} {\bibinfo  {journal}
  {Nature}\ }\textbf {\bibinfo {volume} {407}},\ \bibinfo {pages} {351}
  (\bibinfo {year} {2000})}\BibitemShut {NoStop}%
\bibitem [{\citenamefont {Inosov}\ \emph {et~al.}(2010)\citenamefont {Inosov},
  \citenamefont {Park}, \citenamefont {Bourges}, \citenamefont {Sun},
  \citenamefont {Sidis}, \citenamefont {Schneidewind}, \citenamefont {Hradil},
  \citenamefont {Haug}, \citenamefont {Lin}, \citenamefont {Keimer},\ and\
  \citenamefont {Hinkov}}]{Inosov2010}%
  \BibitemOpen
  \bibfield  {author} {\bibinfo {author} {\bibfnamefont {D.~S.}\ \bibnamefont
  {Inosov}}, \bibinfo {author} {\bibfnamefont {J.~T.}\ \bibnamefont {Park}},
  \bibinfo {author} {\bibfnamefont {P.}~\bibnamefont {Bourges}}, \bibinfo
  {author} {\bibfnamefont {D.~L.}\ \bibnamefont {Sun}}, \bibinfo {author}
  {\bibfnamefont {Y.}~\bibnamefont {Sidis}}, \bibinfo {author} {\bibfnamefont
  {A.}~\bibnamefont {Schneidewind}}, \bibinfo {author} {\bibfnamefont
  {K.}~\bibnamefont {Hradil}}, \bibinfo {author} {\bibfnamefont
  {D.}~\bibnamefont {Haug}}, \bibinfo {author} {\bibfnamefont {C.~T.}\
  \bibnamefont {Lin}}, \bibinfo {author} {\bibfnamefont {B.}~\bibnamefont
  {Keimer}},\ and\ \bibinfo {author} {\bibfnamefont {V.}~\bibnamefont
  {Hinkov}},\ }\bibfield  {title} {\bibinfo {title} {Normal-state spin dynamics
  and temperature-dependent spin-resonance energy in optimally doped
  bafecoas},\ }\href {https://doi.org/10.1038/nphys1483} {\bibfield  {journal}
  {\bibinfo  {journal} {Nature Physics}\ }\textbf {\bibinfo {volume} {6}},\
  \bibinfo {pages} {178} (\bibinfo {year} {2010})}\BibitemShut {NoStop}%
\bibitem [{\citenamefont {Varma}\ \emph {et~al.}(2015)\citenamefont {Varma},
  \citenamefont {Zhu},\ and\ \citenamefont {Schr\"oder}}]{CMVZhuSchroeder2016}%
  \BibitemOpen
  \bibfield  {author} {\bibinfo {author} {\bibfnamefont {C.~M.}\ \bibnamefont
  {Varma}}, \bibinfo {author} {\bibfnamefont {L.}~\bibnamefont {Zhu}},\ and\
  \bibinfo {author} {\bibfnamefont {A.}~\bibnamefont {Schr\"oder}},\ }\bibfield
   {title} {\bibinfo {title} {Quantum critical response function in
  quasi-two-dimensional itinerant antiferromagnets},\ }\href
  {https://doi.org/10.1103/PhysRevB.92.155150} {\bibfield  {journal} {\bibinfo
  {journal} {Phys. Rev. B}\ }\textbf {\bibinfo {volume} {92}},\ \bibinfo
  {pages} {155150} (\bibinfo {year} {2015})}\BibitemShut {NoStop}%
\bibitem [{\citenamefont {Bok}\ \emph {et~al.}(2016)\citenamefont {Bok},
  \citenamefont {Bae}, \citenamefont {Choi}, \citenamefont {Varma},
  \citenamefont {Zhang}, \citenamefont {He}, \citenamefont {Zhang},
  \citenamefont {Yu},\ and\ \citenamefont {Zhou}}]{Bok_ScienceADV}%
  \BibitemOpen
  \bibfield  {author} {\bibinfo {author} {\bibfnamefont {J.~M.}\ \bibnamefont
  {Bok}}, \bibinfo {author} {\bibfnamefont {J.~J.}\ \bibnamefont {Bae}},
  \bibinfo {author} {\bibfnamefont {H.-Y.}\ \bibnamefont {Choi}}, \bibinfo
  {author} {\bibfnamefont {C.~M.}\ \bibnamefont {Varma}}, \bibinfo {author}
  {\bibfnamefont {W.}~\bibnamefont {Zhang}}, \bibinfo {author} {\bibfnamefont
  {J.}~\bibnamefont {He}}, \bibinfo {author} {\bibfnamefont {Y.}~\bibnamefont
  {Zhang}}, \bibinfo {author} {\bibfnamefont {L.}~\bibnamefont {Yu}},\ and\
  \bibinfo {author} {\bibfnamefont {X.~J.}\ \bibnamefont {Zhou}},\ }\bibfield
  {title} {\bibinfo {title} {Quantitative determination of pairing interactions
  for high-temperature superconductivity in cuprates},\ }\href
  {https://doi.org/10.1126/sciadv.1501329} {\bibfield  {journal} {\bibinfo
  {journal} {Science Advances}\ }\textbf {\bibinfo {volume} {2}},\ \bibinfo
  {pages} {e1501329} (\bibinfo {year} {2016})}\BibitemShut {NoStop}%
\bibitem [{\citenamefont {Varma}(2022)}]{Varma_RH2022}%
  \BibitemOpen
  \bibfield  {author} {\bibinfo {author} {\bibfnamefont {C.~M.}\ \bibnamefont
  {Varma}},\ }\bibfield  {title} {\bibinfo {title} {Quantum-critical
  resistivity of strange metals in a magnetic field},\ }\href
  {https://doi.org/10.1103/PhysRevLett.128.206601} {\bibfield  {journal}
  {\bibinfo  {journal} {Phys. Rev. Lett.}\ }\textbf {\bibinfo {volume} {128}},\
  \bibinfo {pages} {206601} (\bibinfo {year} {2022})}\BibitemShut {NoStop}%
\bibitem [{\citenamefont {Son}(1999)}]{son}%
  \BibitemOpen
  \bibfield  {author} {\bibinfo {author} {\bibfnamefont {D.~T.}\ \bibnamefont
  {Son}},\ }\bibfield  {title} {\bibinfo {title} {Superconductivity by
  long-range color magnetic interaction in high-density quark matter},\ }\href
  {https://doi.org/10.1103/PhysRevD.59.094019} {\bibfield  {journal} {\bibinfo
  {journal} {Phys. Rev. D}\ }\textbf {\bibinfo {volume} {59}},\ \bibinfo
  {pages} {094019} (\bibinfo {year} {1999})}\BibitemShut {NoStop}%
\bibitem [{\citenamefont {Miyashita}\ and\ \citenamefont
  {Shiba}(1984)}]{Shiba1984}%
  \BibitemOpen
  \bibfield  {author} {\bibinfo {author} {\bibfnamefont {S.}~\bibnamefont
  {Miyashita}}\ and\ \bibinfo {author} {\bibfnamefont {H.}~\bibnamefont
  {Shiba}},\ }\bibfield  {title} {\bibinfo {title} {Nature of the phase
  transition of the two-dimensional antiferromagnetic plane rotor model on the
  triangular lattice},\ }\href@noop {} {\bibfield  {journal} {\bibinfo
  {journal} {J. Phys. Soc. Japan}\ }\textbf {\bibinfo {volume} {53}},\ \bibinfo
  {pages} {1145} (\bibinfo {year} {1984})}\BibitemShut {NoStop}%
\bibitem [{\citenamefont {Kohn}\ and\ \citenamefont {Luttinger}(1965)}]{KL}%
  \BibitemOpen
  \bibfield  {author} {\bibinfo {author} {\bibfnamefont {W.}~\bibnamefont
  {Kohn}}\ and\ \bibinfo {author} {\bibfnamefont {J.~M.}\ \bibnamefont
  {Luttinger}},\ }\bibfield  {title} {\bibinfo {title} {New mechanism for
  superconductivity},\ }\href {https://doi.org/10.1103/PhysRevLett.15.524}
  {\bibfield  {journal} {\bibinfo  {journal} {Phys. Rev. Lett.}\ }\textbf
  {\bibinfo {volume} {15}},\ \bibinfo {pages} {524} (\bibinfo {year}
  {1965})}\BibitemShut {NoStop}%
\bibitem [{\citenamefont {Raghu}\ \emph {et~al.}(2010)\citenamefont {Raghu},
  \citenamefont {Kivelson},\ and\ \citenamefont {Scalapino}}]{Raghu2010}%
  \BibitemOpen
  \bibfield  {author} {\bibinfo {author} {\bibfnamefont {S.}~\bibnamefont
  {Raghu}}, \bibinfo {author} {\bibfnamefont {S.~A.}\ \bibnamefont
  {Kivelson}},\ and\ \bibinfo {author} {\bibfnamefont {D.~J.}\ \bibnamefont
  {Scalapino}},\ }\bibfield  {title} {\bibinfo {title} {Superconductivity in
  the repulsive hubbard model: An asymptotically exact weak-coupling
  solution},\ }\href {https://doi.org/10.1103/PhysRevB.81.224505} {\bibfield
  {journal} {\bibinfo  {journal} {Phys. Rev. B}\ }\textbf {\bibinfo {volume}
  {81}},\ \bibinfo {pages} {224505} (\bibinfo {year} {2010})}\BibitemShut
  {NoStop}%
\bibitem [{\citenamefont {Baranov}\ \emph {et~al.}(1992)\citenamefont
  {Baranov}, \citenamefont {Chubukov},\ and\ \citenamefont {Kagan}}]{kagan}%
  \BibitemOpen
  \bibfield  {author} {\bibinfo {author} {\bibfnamefont {M.~A.}\ \bibnamefont
  {Baranov}}, \bibinfo {author} {\bibfnamefont {A.~V.}\ \bibnamefont
  {Chubukov}},\ and\ \bibinfo {author} {\bibfnamefont {M.~Y.}\ \bibnamefont
  {Kagan}},\ }\bibfield  {title} {\bibinfo {title} {Superconductivity and
  superfluidity in fermi systems with repulsive interactions},\ }\href
  {https://doi.org/10.1142/S0217979292001249} {\bibfield  {journal} {\bibinfo
  {journal} {International Journal of Modern Physics B}\ }\textbf {\bibinfo
  {volume} {06}},\ \bibinfo {pages} {2471} (\bibinfo {year}
  {1992})}\BibitemShut {NoStop}%
\bibitem [{\citenamefont {Chubukov}(1993)}]{Chubukov93}%
  \BibitemOpen
  \bibfield  {author} {\bibinfo {author} {\bibfnamefont {A.~V.}\ \bibnamefont
  {Chubukov}},\ }\bibfield  {title} {\bibinfo {title} {Kohn-luttinger effect
  and the instability of a two-dimensional repulsive fermi liquid at t=0},\
  }\href {https://doi.org/10.1103/PhysRevB.48.1097} {\bibfield  {journal}
  {\bibinfo  {journal} {Phys. Rev. B}\ }\textbf {\bibinfo {volume} {48}},\
  \bibinfo {pages} {1097} (\bibinfo {year} {1993})}\BibitemShut {NoStop}%
\bibitem [{\citenamefont {Long}\ \emph {et~al.}(2024)\citenamefont {Long},
  \citenamefont {Jimeno-Pozo}, \citenamefont {Sainz-Cruz}, \citenamefont
  {Pantalean},\ and\ \citenamefont {Guinea}}]{Guinea2024}%
  \BibitemOpen
  \bibfield  {author} {\bibinfo {author} {\bibfnamefont {M.}~\bibnamefont
  {Long}}, \bibinfo {author} {\bibfnamefont {A.}~\bibnamefont {Jimeno-Pozo}},
  \bibinfo {author} {\bibfnamefont {H.}~\bibnamefont {Sainz-Cruz}}, \bibinfo
  {author} {\bibfnamefont {P.~A.}\ \bibnamefont {Pantalean}},\ and\ \bibinfo
  {author} {\bibfnamefont {F.}~\bibnamefont {Guinea}},\ }\bibfield  {title}
  {\bibinfo {title} {Evolution of superconductivity in twisted graphene
  multilayers},\ }\href@noop {} {\bibfield  {journal} {\bibinfo  {journal}
  {Proceedings of the National Academy of Sciences}\ }\textbf {\bibinfo
  {volume} {121}},\ \bibinfo {pages} {e2405259121} (\bibinfo {year}
  {2024})}\BibitemShut {NoStop}%
\end{thebibliography}%
\end{document}